# Are Work Zones and Connected Automated Vehicles Ready for a Harmonious Coexistence? A Scoping Review and Research Agenda

**Amjad Dehman, Ph.D., Senior Researcher***
Laboratory of Innovations in Transportation (Litrans)
Centre for Urban Innovation, Ryerson University
44 Gerrard Street East, Toronto ON M5G 1G3, Canada
Emails: amjad.dehman@ryerson.ca, amjad.dehman@gmail.com
Phone: 416-979-5000 ext. 556456

**Bilal Farooq, Ph.D., Associate Professor**
Laboratory of Innovations in Transportation (Litrans)
Centre for Urban Innovation, Ryerson University
44 Gerrard Street East, Toronto ON M5G 1G3, Canada
Emails: bilal.farooq@ryerson.ca
Phone: 416-979-5000 ext. 556456

* Corresponding Author.
Declaration of Interest: None

**Abstract**

The recent advent of connected and automated vehicles (CAVs) is expected to transform the transportation system. CAV technologies are being developed rapidly and they are foreseen to penetrate the market at a rapid pace. On the other hand, work zones (WZs) have become common areas on highway systems as a result of the increasing construction and maintenance activities. The near future will therefore bring the coexistence of CAVs and WZs which makes their interaction inevitable. WZs expose all vehicles to a sudden and complex geometric change in the roadway environment, something that may challenge many of CAV navigation capabilities. WZs however also impose a space contraction resulting in adverse traffic impacts, something that legitimately calls for benefiting from the highly efficient CAV functions. CAVs should be able to reliably traverse WZ geometry and WZs should benefit from CAV intelligent functions. This paper reviews the state-of-the-art and the key concepts, opportunities, and challenges of deploying CAV systems at WZs. The reviewed subjects include traffic performance and behaviour, technologies and infrastructure, and regulatory considerations. Eighteen CAV mobility, safety, and environmental concepts and functions were distributed over the WZ area which was subdivided into five segments: further upstream, approach area, queuing area, WZ activity, and termination area. In addition, among other topics reviewed and discussed are detection of WZ features, smart traffic control devices, various technologies at connected WZs, cross-border harmonization, liability, insurance, and privacy. The paper also provides a research agenda with a list of research needs supported by experts rating and inputs. The paper aims to provide a bird's eye view, but with necessary details that can benefit researchers, practitioners, and transportation agencies.

**Keywords:** Connected and automated vehicles, work zone, traffic safety, environment, traffic control devices, traffic regulations.

1. **INTRODUCTION**

Connected and Automated Vehicles (CAVs) are being developed rapidly and they are foreseen to considerably penetrate the market in the near future. Their advanced and smart functions can remarkably improve highway mobility and safety. On the other hand, work zones (WZs) have become common areas on the highway system in response to the growing construction and maintenance activities. The coexistence of CAVs and WZs has so become inevitable. The interesting question would be *"Is this coexistence an opportunity, a challenge, or both?"* WZs may benefit from CAV smart functions to improve traffic flow performance, CAVs however need to accurately perceive the complex geometry of WZs and traverse these areas reliably.

Statistics have been increasingly and continuously stressing on the impact of work zones on mobility and safety. The US Federal Highway Administration (FHWA) [1] estimated that construction causes around 10% of the total road delay in the U.S., which comes after recurring bottlenecks (40%), traffic incidents (25%), and inclement weather (15%). This 10% work-zone-induced congestion share can cause an enormous financial impact knowing that the overall traffic congestion cost is substantially high. A study by Metrolinx and HDR Corporation [2] estimated that the cost of congestion in the Greater Toronto and Hamilton Area is around $6 billion annually based on the 2006 travel data, and this figure is forecasted to increase to $15 billion per year by 2031. The annual cost of traffic congestion in the US was estimated to be around $179 billion based on 2017 travel data covering 494 U.S. urban areas [3]. Beside mobility impacts, WZs also create significant safety challenges. According to FHWA statistics, work zones resulted in 587 traffic-related fatalities in 2011 and 37,000 injuries in 2010; also, an average of two people are killed and 101 are injured every day in U.S. highway work zones [4]. The adverse WZ-induced mobility and safety impacts legitimately call for investing heavily in advanced technologies and smart systems. With the looming CAV penetration in the highway network, benefiting from the highly efficient CAV mobility and safety functions at WZs is expected to be an active and important field of research.

Using the Transport Research International Documentation (TRID), an integrated database that combines more than 1.25 million records of transportation-related research worldwide, Figure 1 demonstrates the evolution of scientific publications relevant to both CAVs and WZs in the last three decades. The search keywords used were "work zone" and "connected and automated vehicles." The surveyed documents were not limited to journal articles but also included conference papers, technical reports, and abstracts of active and recent projects. For WZ-publications, the average number of annual publications experienced a remarkable climb in the 2000-2009 period as compared to the 1990's, i.e., 179.9 versus 97.5 publications/year. WZ-publications continued to grow moderately in the last decade reaching 196.2 publications/year in the 2010-2019 period. On the other hand, only a handful of CAV-publications were noticed, if any, prior to 2010 after which the number of CAV-publications has been growing rapidly and even surpassing WZ-publications. However, when using the term "connected and automated vehicles at work zones," only 12 records were retrieved, and these were in the period 2015-to-2019. Thus, deploying CAV systems at WZs is still at its early emergence and yet plenty of publications are forthcoming as a natural consequence of the increasing trend in both WZ- and CAV-publications.

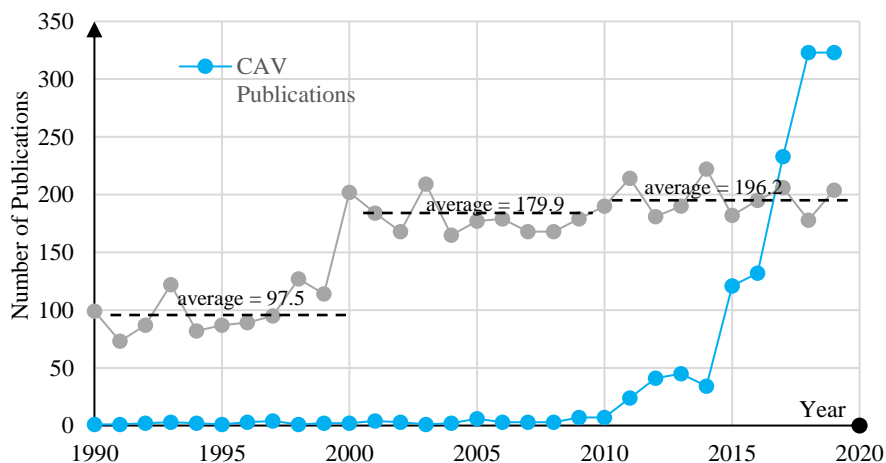

Notes: numbers were based on TRID database as of October 28, 2020, the covered period is from 1990-2019

**Figure 1. Evolution of CAV and WZ publications in the last three decades.**

The previous discussion highlights two main points; first, the high importance of integrating or deploying CAVs at WZs, and second, the fact that this field of research is still in its early emergence. There is therefore a need for a paper that establishes a framework for this research direction and paves the way forward. This paper responds to this motivation by:

*(i)* *investigating how CAVs and WZs will interact and discussing the resulting key concepts, opportunities, and challenges,*

*(ii)* *reviewing the state-of-the-art and two categories of the literature, i.e., studies that particularly explored CAV concepts at WZs albeit only few of such efforts already exist and studies that have relevant and transferable concepts which can be applied at WZs, and*

*(iii)* *presenting a research agenda and a list of research needs supported by experts' rating and inputs.*

The aim is to provide a bird's eye view along with supporting details in order to map the deployment of CAVs at WZs and to benefit practitioners, planners, transportation agencies, and prospective researchers who are interested in understanding the future of WZs.

The paper is organized as shown in Figure 2. After this introduction, Sections 2, 3, and 4 constitute the state-of-the-art review of CAV systems at WZs, they respectively represent traffic performance and behaviour, technologies and infrastructure, and finally regulatory considerations. In Section 5, a research framework is outlined and research needs are listed. A conclusion summary is finally provided in Section 6.

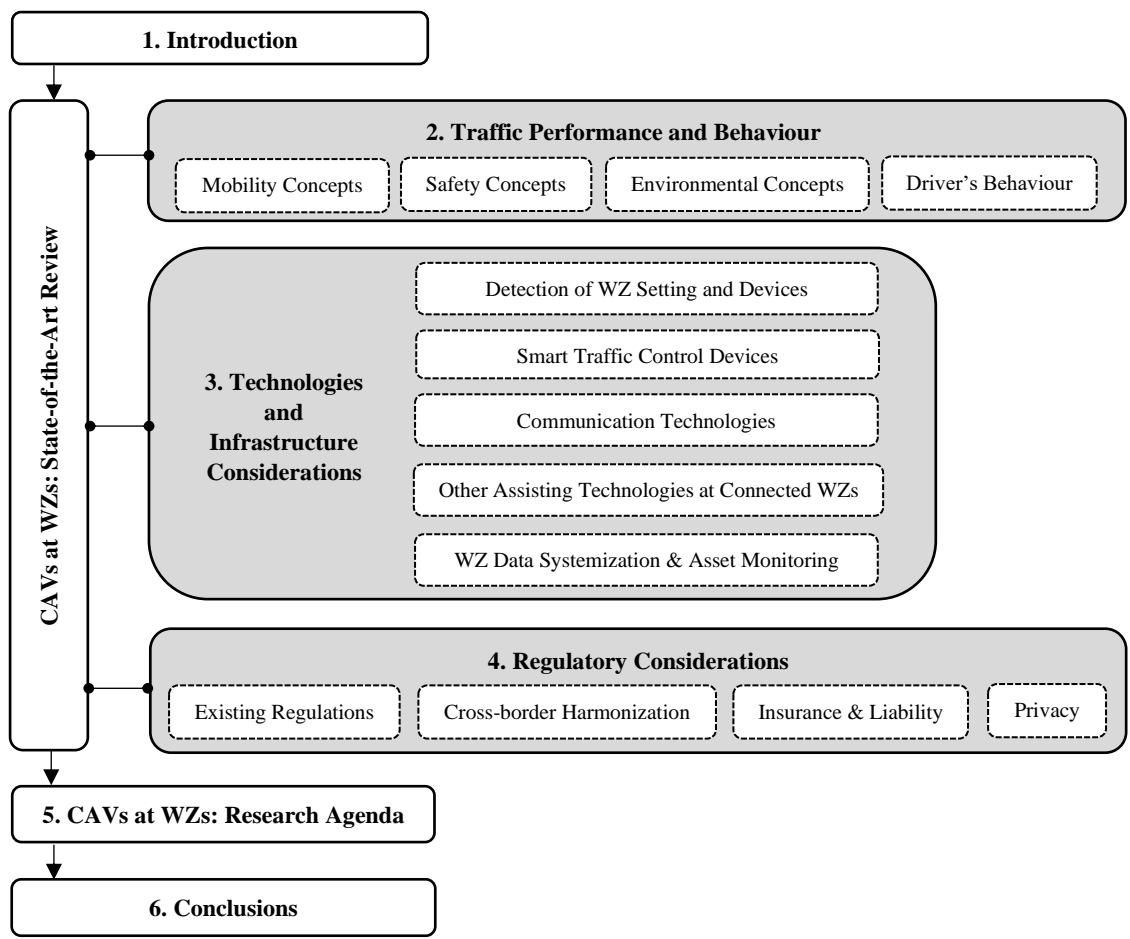

**Figure 2. The paper roadmap.**

## 2. TRAFFIC PERFORMANCE AND BEHAVIOUR

The potential contribution of CAVs to improve traffic performance at WZs is discussed in this section from three perspectives, mobility, safety, and environment. The aim is to enhance the efficiency of traffic mobility while also lowering crash rates, fuel consumption, and emissions. Several CAV concepts and functions that benefit each of these perspectives are discussed and reviewed, some concepts may appear more than one time if they benefit several perspectives at the same time. Automation is different from connectivity, vehicle functions may rely on one of them or both. However, this paper generally uses the abbreviation CAVs combining automation and connectivity without detailing the level and contribution size of each. Nevertheless, in the following discussed concepts, the title of each indicates in parentheses whether the concept relies on connectivity, automation, or both technologies. This is indicated from a high-level perspective, so if a function is mainly "connectivity-based" this does not mean it cannot also benefit from automation and vice versa.

### 2.1 Mobility Concepts and Functions

Mobility at WZs refers to traffic operational measures such as capacity, demand, level-of-service, delay, and travel time. The objective is to optimize WZ capacity and reduce or at least manage the upstream demand so the traffic can traverse the WZ smoothly and quickly. In the following sections, the discussed CAV mobility concepts were ordered in a way that the basic or universal concepts come first and those concepts that rely on others follow them.

*2.1.1 Tightened Following Gap (Automation- and Connectivity-Based)*

CAVs are expected to shorten the following gap allowing better utilization of the highway space. Nowakowski et al. [5] indicated that CAVs supported by V2V communication can reduce the mean following time of human-driven vehicles from 1.40 sec to 0.60 sec. A shorter following gap means higher capacity which is vitally needed at congested WZs to reduce the chance of breakdown or to discharge the queues more efficiently. Most adaptive cruise control (ACC) systems follow a "speed-based gap" whereby the distance between vehicles is proportional to their speed plus a fixed standstill distance [6]. This is especially important at WZ, where the traffic speed fluctuates widely. The tightened gap between CAVs is mainly based on their ability to accelerate and decelerate assertively and promptly. They use connectivity to obtain data of high-accuracy and low-latency of the lead vehicle motion and use automation to eliminate the imperfection in driver's perception-reaction [7, 8].

*2.1.2 CAV Clustering (Mainly Connectivity-Based)*

If CAVs are distributed randomly in a mixed traffic stream, the chance of CAV-coupling will only depend on the market penetration rate (MPR). More CAV-generated short-gaps can be realized if CAVs can deliberately follow each other. CAV clustering instructs equipped-vehicles that are travelling in close proximity to form platoons and strings. To do so, CAVs may have to accelerate, decelerate, and change lanes. They have also to be provided with accurate data about the motion of nearby equipped-vehicles including their travel lane. Associating nearby CAVs with their exact travel lane can be challenging at WZs having complex and varying lane configurations. This warrants the use of special and smart traffic control devices as discussed later in *Sections 3.1 and 3.2*. Shladover et al. [6] defined "Global Clustering" when vehicles sharing the same destination meet early in the network to form platoons, the authors however considered this challenging because it requires wide communication coverage and waiting areas for vehicles to meet.

Although clustering is generally beneficial for congested traffic, it may result in adverse impacts on some lanes near the WZ taper. Figure 3 illustrates some possible impacts of clustering by different lanes. Clustering on the closed lane creates long platoons that may not be able to find sufficiently wide merging gaps on the adjacent lane (Figure 3a). If traffic demand is persistently high on the adjacent open lanes, dissolution of the stranded clusters may become inevitable. Additionally, if the rate of clustered vehicles is excessively high on the first open lane (Figure 3b), long and dense platoons will be passing repeatedly on that lane. Consequently, vehicles on the closed lane will not find merging gaps easily and they may have to endure excessive queuing delay. This problem can especially impact human-driven vehicles that cannot communicate with CAVs and ask for cooperation. Alternatively, clustering may be better limited to the far open lanes at high traffic demand conditions (Figure 3c). This ensures easier lane change maneuvers from the closed lane to the first open lane for both the human-driven vehicles and the non-clustered CAVs. Another advanced option is to assign a dedicated lane for CAVs (Figures 3d and 2e) that is further discussed in the following concept (*Section 2.1.3*). Permitting or prohibiting clustering on specific lanes and how this can be influenced by WZ lane configurations (2-to-1, 3-to-1, 4-to-1, 4-to-2, etc), traffic demand level, MPR, maximum platoon (cluster) length, and the distance remaining between the approaching clusters and the WZ taper are critical topics that merit a detailed investigation.

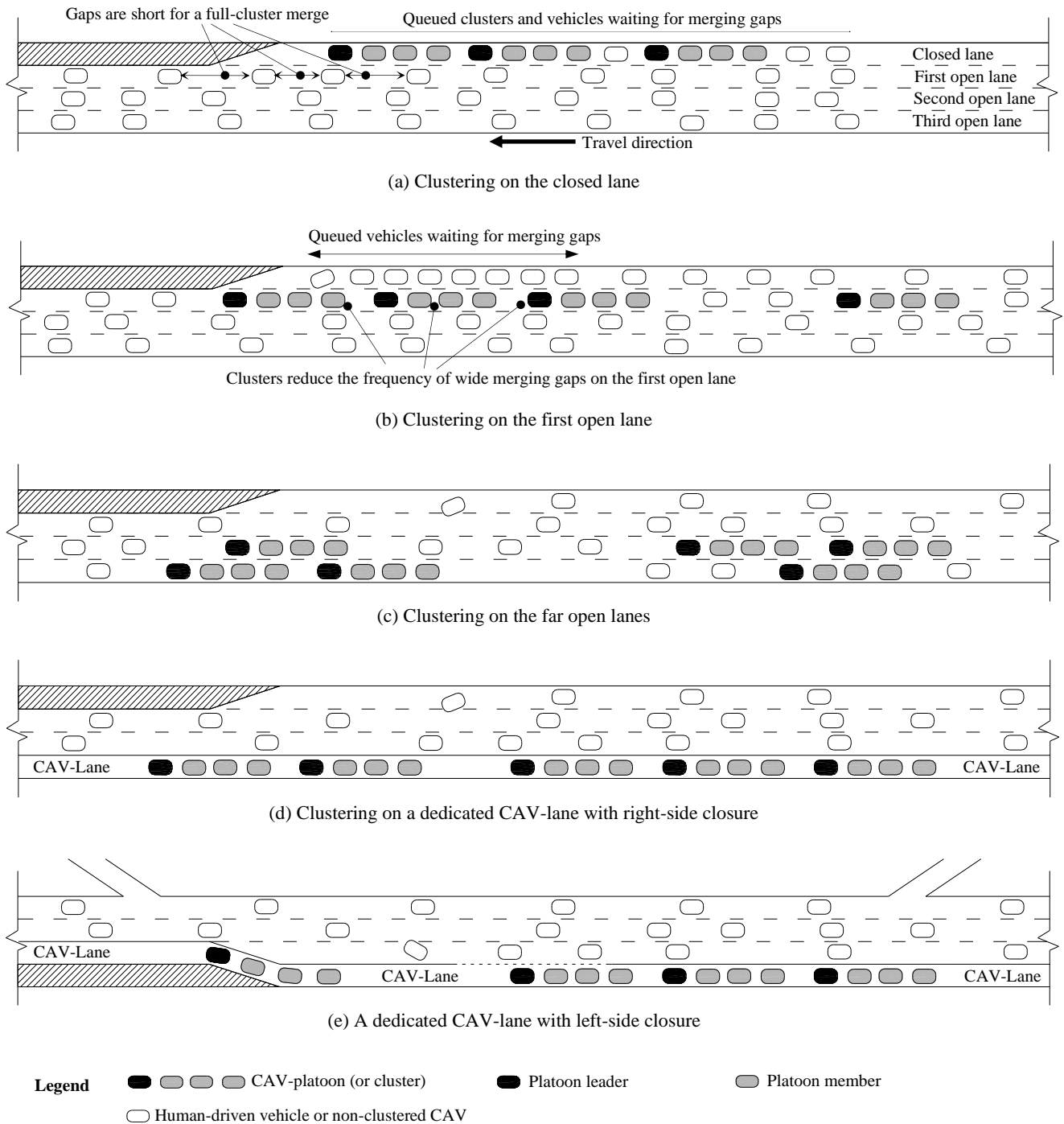

**Notes:** distances are not to scale and for illustration purpose only, maximum platoon length is assumed to be four vehicles

**Figure 3. CAV clustering by different lanes in the WZ vicinity.**

### *2.1.3 Dedicated CAV Lane (Automation- and Connectivity-Based)*

An advanced level of clustering CAVs would be to assign them a dedicated lane, this lane however can still accept human-driven vehicles if the MPR is not high enough. A dedicated lane would mainly improve traffic efficiency and increase capacity by utilizing the smart car-following between CAVs while also promising for cluster-based safety and environmental benefits. The capacity of a dedicated CAV-lane can reach 3400 veh/hr/ln at 90% MPR [*9*]. Moreover, CAV-lane allows for more stable communication density which makes the communication more reliable with low chance of packet drop [*9*]. Some studies estimated that a minimum of 20% MPR in the network and a four-lane basic section are needed to justify a dedicated CAV-lane [*10, 11*]. A dedicated CAV-lane should be typically assigned on the furthest left lane in order to avoid conflicting with ramp junctions. This perfectly suits a right-side closure (Figure 3d). If the closure is on the left-side, a buffering taper is needed to maintain the CAV-lane continuity and transfer the closure to the other mixed-traffic lanes (Figure 3e).

### *2.1.4 Synchronized Acceleration/Deceleration (Automation- and Connectivity-Based)*

When human-driven vehicles discharge from a queue, vehicles in front cannot move simultaneously and they must accelerate and discharge one after the other resulting in a start-up lost time. The V2V-powered following behaviour allows a platoon of vehicles to accelerate synchronously in a train or semi-train fashion eliminating or reducing the start-up lost time. With V2V communication, the fourth following vehicle can be notified of the lead vehicle action within 400 to 800 ms if the data are propagated pair-wise, or within only 100 ms if the data are propagated platoon-wise upstream irrelevant to the line of sight [*6*]. The synchronized acceleration/deceleration can fit well in congested WZs where vehicles discharge from queued traffic. With a traffic signal, a V2I communication between queued vehicles and the signal can make the movement of a train of vehicles perfectly synchronized with no loss in the green phase time. It is worth noting that traffic signals are not only suitable at arterial WZs, but may also be deployed at freeway WZs as supported by recent studies. Ramadan et al. [*12*] and Tympakianaki et al. [*13*] evaluated the use of a traffic signal on the freeway mainline immediately upstream the WZ in order to reduce the lane change friction between vehicles and to provide an organized queue discharge discipline. The studies indicated favourable safety and mobility outcomes during peak-hour conditions.

### *2.1.5 Lane Merge Control (Automation- and Connectivity-Based)*

Lane merge control at WZs is a well-established research subject and several merge strategies have been proposed and evaluated. Figure 4 demonstrates three different merge control strategies at WZs. In the late-merge (Figure 4a), drivers are advised to stay on their lane, they are then instructed to follow a zipper-merge control just immediately before the WZ taper. This approach can be disadvantageous to CAV operations since it results in splitting CAV clusters that were formed upstream. In the early-merge (Figure 4b), drivers are advised to change lane early upstream to avoid last-moment aggressive merging. This can protect CAV clusters, especially if "CAV cooperative lane change" is deployed upstream to enhance the clustering. However, early-merge control does not perform well at congested sites, it shifts the queues further upstream, and in extreme cases, the early-merge will almost operate like a late-merge system with a merging point shifted upstream. With a traffic signal control (Figure 4c), vehicles occupy all lanes and follow a phasing scheme. A full-traffic-cycle instead of one-car-per-green is deployed and each lane is provided with a distinct traffic signal, so the lanes do not discharge traffic simultaneously. This approach protects CAV clusters and allows for a platoon-wide synchronized acceleration using the V2I communication between CAVs and the signal controller. The phasing can be fixed allowing for (n) vehicles to discharge per green or can be actuated based on the available clusters, i.e., the green time may be slightly extended or shortened to allow clustered vehicles to discharge at once. A detailed comparison between the three merge control strategies is recommended as a future research.

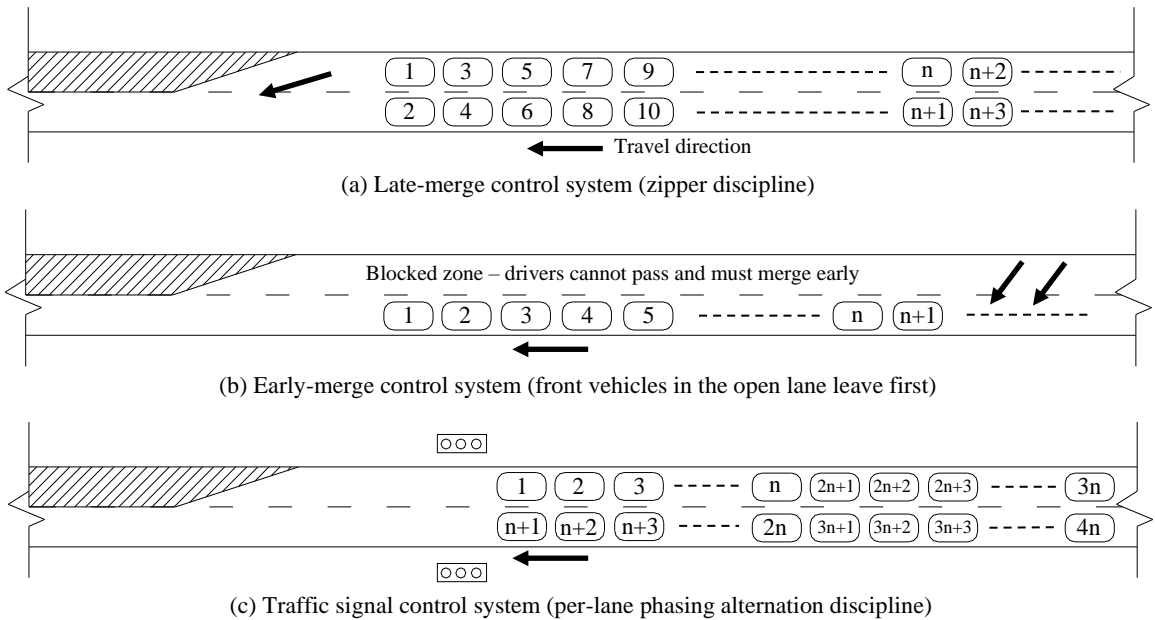

(a) Late-merge control system (zipper discipline)

(b) Early-merge control system (front vehicles in the open lane leave first)

(c) Traffic signal control system (per-lane phasing alternation discipline)

**Note:** the numbers refer to the order of the discharged vehicles

**Figure 4. Different lane merge control systems.**

*2.1.6   Cooperative Lane Change (Mainly Connectivity-Based)*

The cooperative lane change fits well with the closure-based WZ areas, where the lane change becomes mandatory for all vehicles travelling on the closed lanes. The most common cooperative lane change concept typically involves a pair of vehicles as shown in Figure 5a. The first vehicle (Vehicle A) signals to nearby vehicles explicitly that it needs to change its lane. The second vehicle (Vehicle B), which is travelling on the target lane immediately upstream the first vehicle, will recognize the signal and then decelerate to allow for a smooth and reliable lane change. Vehicles behind Vehicle B may also need to decelerate and cooperate if travelling closely. Vehicle B on the target lane may also collaborate by switching to the next adjacent lane so it creates an adequate merging gap for vehicle A (Figure 5b). When traffic demand increases, the decision of which vehicle needs to collaborate with which vehicle becomes more complex as shown in Figure 5c. If vehicles A1 and A2 signal their need to change lane, vehicle B1 may decide that it is too late to cooperate and it therefore speeds up and moves forward. Then, vehicle B2 is left with a complex decision, i.e., if it decides to cooperate, it needs to perceive the projected trajectories and speeds of both vehicles A1 and A2 and decide accordingly which one is more feasible to collaborate with. Cooperative lane change for such complex traffic flow situations can be addressed by incorporating sophisticated algorithms [*14*]. Alternatively, Ren et al. [*15*] proposed a strategy whereby vehicles on the open lane are instructed to maintain a minimum distance from each other (i.e., equal to two times the safe distance) early upstream of the WZ. When the WZ taper becomes close, vehicles on the closed lane can subsequently merge easily to the open lane as in Figure 5d. The applicability of this approach however requires 100% MPR.

Platoon-based lane change is an under-researched subject and can be challenging at WZs because of their confined space. Hsu et al. [*16*], for example, proposed different platoon-specific lane change algorithms based on the communication time; however, the impact of different traffic flow levels was not examined. Bevly et al. [*17*] reviewed merge and lane change control methods and their algorithms, the study called for more efforts that explore platoon-based lane change methods.

V2V communication should be used for all cooperative concepts because existing V2I communication systems may not transmit data with sufficient latency and accuracy, the V2V communication delay however must still be carefully considered [*18*]. Research is recommended to elaborate more on developing cooperative lane change algorithms and strategies that consider different MPRs, urgency of the lane change request, the distance remaining to the WZ taper, communication latency, WZ lane configuration, traffic demand level, and the possibility of cooperating with human-driven vehicles.

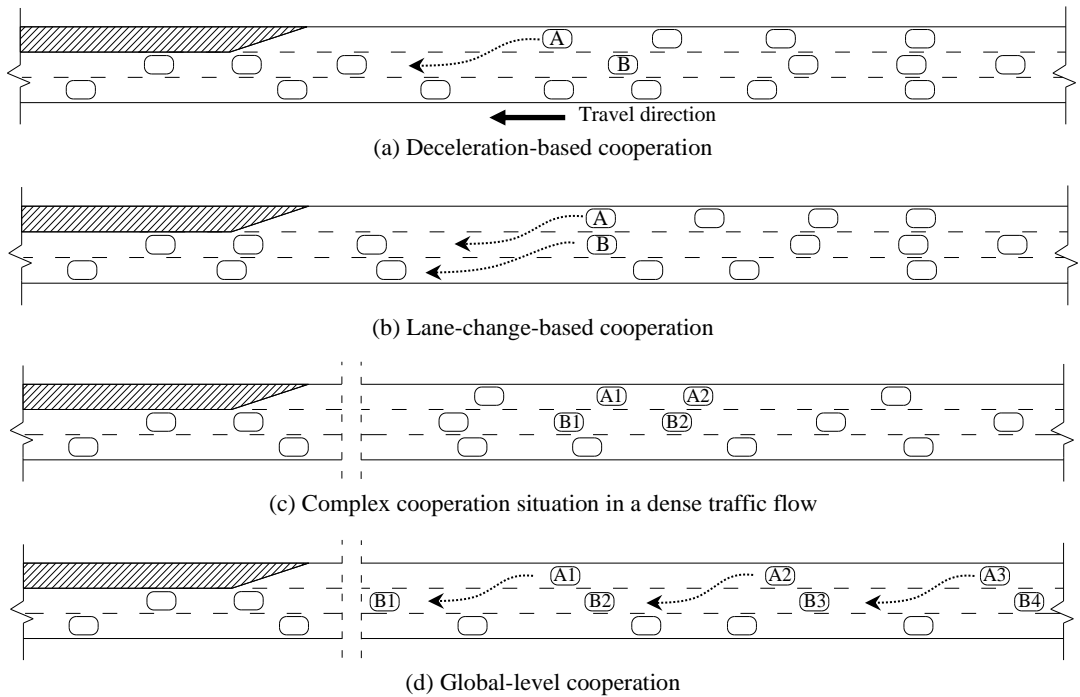

**Figure 5. Cooperative lane change scenarios.**

### 2.1.7 Smart Lane Flow Distribution (Mainly Connectivity-Based)

Achieving a balanced flow distribution over the traffic lanes is expected to optimize capacity and enhance safety. A freeway section usually has higher flow on the fastest left-side lanes. This generates a heavy demand for lane changes at a WZ with left-side closure. CAV-powered smart lane flow distribution depends on early and gradual lane assignment strategies whereby the system randomly selects a group of drivers to change lane, then another group and so forth until a desired lane balance is achieved. Drivers who still have not changed lane from the preceding group are given priority, the number of simultaneous lane changes must be limited. Schakel et al. [*7*] used V2I communication to advise drivers to change lanes aiming to achieve an optimal lane flow distribution. The system was applied at several highway configurations including a lane-drop case; it delayed the onset of breakdown and shortened the overall travel time by 49%.

### 2.1.8 Early Rerouting (Mainly Connectivity-Based)

Rerouting schemes are common traffic management applications at WZs, they advise drivers of available alternate routes. While this initially reduces average trip travel time, it also results in lowering traffic demand at the WZ. The conventional non-CAV rerouting strategies in use today suffer from a small divergence rate because the drivers feel that they are still missing detailed, accurate, and updated information [*19-21*]. CAV-based rerouting at WZs can be smarter in different ways. First, the wide communication coverage of CAV systems allows drivers to alter their routes early at the onset of their trips based on their origin-destination, they do not have to wait until seeing advisory signs upstream the WZ area. Second, a multitude of rerouting options can be provided. Third, traffic conditions on the alternate routes will be updated more frequently and transmitted with low latency allowing for timelier decisions.

Figure 6 outlines the spatial distribution of the previously discussed mobility functions and concepts at WZs, the figure also shows safety and environmental functions which are to be discussed in the following sections. The WZ area is divided into five sub-areas each representing specific traffic conditions and driver tasks. The range of each function denotes where it is most needed and effective. For example, adaptive cruise control is very effective at the queuing area until discharging from the termination area, it can still however be used early upstream if desired by the driver.

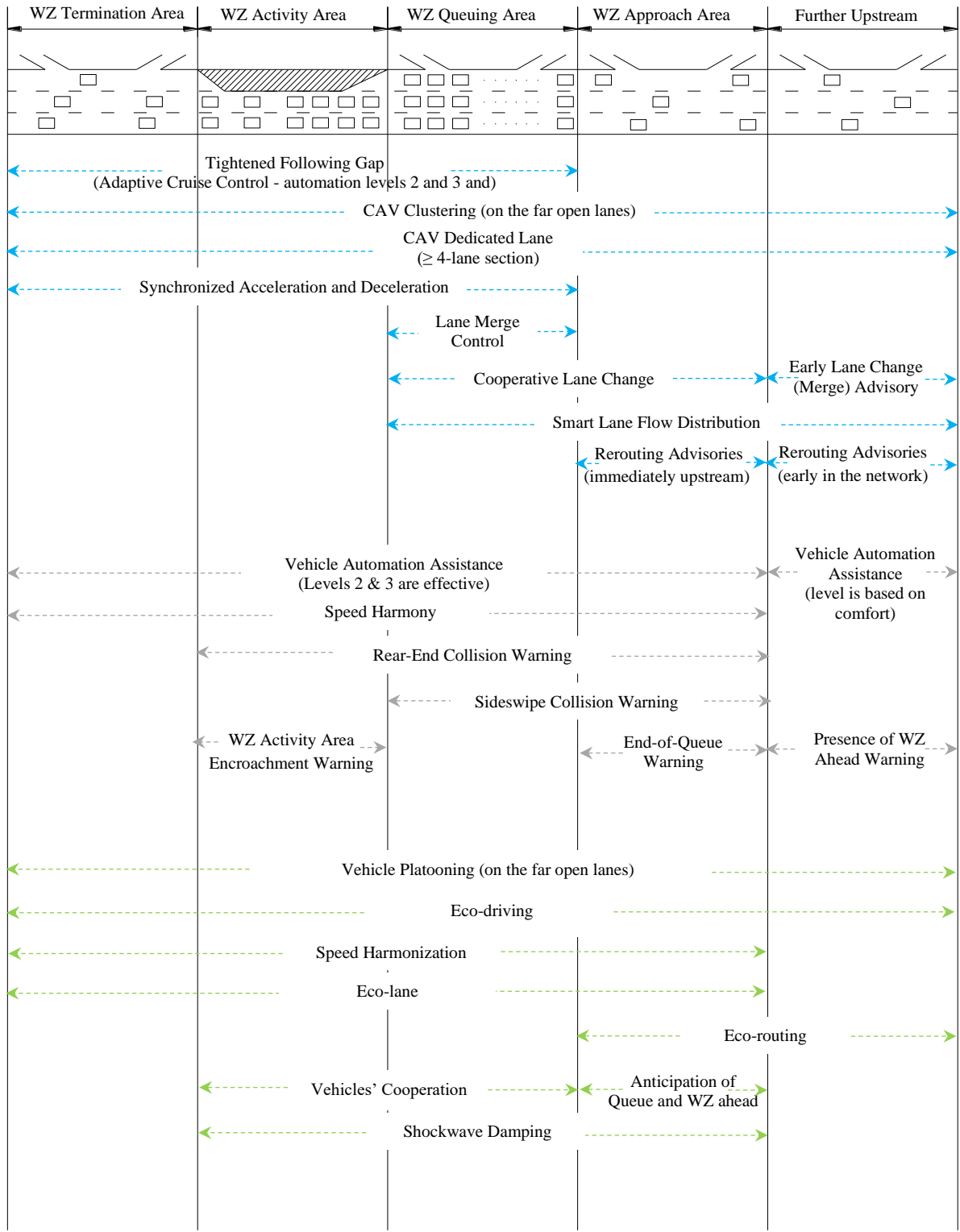

**Note:** blue, gray, and green arrows refer to mobility, safety, and environmental functions respectively.

**Figure 6. Spatial distribution of CAV mobility, safety, and environmental functions at WZ areas.**

**2.2 Safety Concepts and Functions**

Safety at WZs refers herein to reducing or eliminating the chance of accidents by relying on automation and connectivity. Vehicle automation is discussed first because of its universal role in eliminating drivers' errors. Subsequently, connectivity-based concepts specific to WZ vicinities are introduced. The spatial distribution of safety functions and concepts at WZ areas was outlined in Figure 6 ahead.

*2.2.1 Vehicle Automation (Mainly Automation-Based)*

Several studies have indicated that human errors are the major causes of WZ crashes [*22-24*]. Chambless et al. [*22*] estimated that human factors contribute to around 82.7% of all WZ crashes. The high proportion of driver-related crashes makes vehicle automation a promising and vital countermeasure. When automation carries out the driving tasks, driver errors and imperfections will be eliminated. In an optimistic estimation, all crashes caused by the driver-related reasons can be avoided. Automation benefits however will be restricted by some factors, e.g., MPR, automation level, supplying the WZ area with automation-needed infrastructure, and the automation malfunction or error rates if likely. The average levels of automation (Levels 1 to 3) will penetrate the market heavily in the short-run, these levels should allow for adaptive cruise control, which is suitable for common WZ conditions like jam and stop-and-go conditions.

*2.2.2 Speed Harmonization (Mainly Connectivity-Based)*

All vehicles have to frequently change their speed when traversing a WZ depending on whether they are approaching the WZ, traversing the work activity area, or discharging downstream. When congestion prevails, the stop-and-go regime adds extra speed oscillation. Speed variability constitutes a major safety hazard. Speed harmonization is a countermeasure technique that adjusts the advisory speed dynamically based on real-time data. Ma et al. [*25*] reviewed the literature on speed harmonization methods using connectivity, automation, or both. The review showed that a wide range of communication types can be used (V2I, V2V, and hybrid). A simple speed harmonizer can be a V2I system that measures the speed within the WZ using a roadside unit (RSU) and transmits advisory messages to approaching drivers. For WZs with a long impact area upstream, the highway can be subdivided into several segments each having its own RSU that measures the local speed and advises the equipped-vehicles accordingly. Recent studies [*e.g., 26, 27*] showed that CAV-powered speed harmonization can be useful even with low MPRs.

*2.2.3 Crash Warnings (Mainly Connectivity-Based)*

The existing literature mostly agreed that rear-end collision is the most predominant type of WZ crash and that other common types include sideswipe, angle, and hit-fixed-object collisions [*24, 28, 29*]. These crashes also follow important spatial trends [*28*]. Rear-end collision is the most prevailing type everywhere along the WZ area; however, the frequency of sideswipe collision critically increases in the transition area, and the frequencies of the angle and hit-fixed-object collisions critically increase in the work activity area. The prevalence of common crash types at WZs and their spatial trends suggest establishing specific CAV-powered crash warnings as follows.

First: Rear-End Warning

A V2V communication should be established so that the following vehicle can receive from the leading vehicle its position, speed, and acceleration. When a crash becomes imminent, a standard warning message can be triggered and transmitted to the following (striking) vehicle.

Second: Sideswipe and Angle Collision Warning

Warning for sideswipe and angle collisions needs the information on all vehicles travelling in adjacent lanes. This warrants accurate V2V communication that can detect the exact lane of nearby vehicles while also measuring their motion features.

Third: Work-Activity Area Encroachment Warning

Workers can have wearable connectivity devices that track their position and activity (*Section 3.4.2*). Similarly, traffic control devices (e.g., cones and barrels) and construction equipment can be equipped with smart devices that disclose their identity and location (*Section 3.2.1*). Using V2I (RSU) or V2X (direct communication), approaching vehicles can be warned when becoming too close to workers, devices, or equipment, the workers can also similarly be warned.

All crash warnings discussed above are primarily dependent on V2V and V2X connectivity; however, vehicle automation would also enhance the perception-reaction time to any warning message. In the absence of V2V communications, a warning message can still be transmitted to the striking vehicle with only V2I communication, but the struck vehicle should be adequately equipped and connected to the RSU so it can disclose its position and motion characteristics. Nevertheless, V2I communication will be restricted by the RSU range and the limited latency may adversely impact the timeliness of the warning message.

### *2.2.4 WZ Presence and Queue Warning (Mainly Connectivity-Based)*

Warning drivers of a WZ or end-of-queue ahead can be typically done with V2I systems. The RSU can measure and update the queue length by tracking the speed of the connected vehicles when they pass the WZ area. This eliminates the need for traditional traffic detectors and their associated cost which can be especially impractical for short-term and mobile WZs. V2I communication is usually sufficient, but the V2V communication can increase coverage range when only one RSU is available and solve for excessively long and rapidly changed queues.

## 2.3 Environmental Concepts and Functions

As the attention towards fuel efficiency and emissions increased recently, this section discusses the environmental impact of CAVs at WZs and the potential of these vehicles to reduce emissions and fuel consumptions. Several CAV environmental concepts and functions at WZs are discussed below, these were distributed spatially in Figure 6 ahead.

### *2.3.1 Eco-Driving (Automation- and Connectivity-Based)*

Eco-driving resolves for the imperfection and aggressiveness associated with human driving, it relies on technologies to moderate cruise speed, use gentle acceleration or deceleration rates, avoid abrupt manoeuvres, and keep the vehicle constantly predictive and vigilant about the roadway and traffic changes. Brown et al. [*30*] and Wadud et al. [*31*] estimated an overall reduction of 15% and 20% respectively in fuel consumption due to eco-driving. Gonder et al. [*32*] found that efficient driving behaviour can reduce fuel consumption by 20% for aggressively driven trips and by 5-10% for moderate driving styles. Wu et al. [*33*] tested a fuel-economy system that sends real-time optimal acceleration and deceleration values to drivers, the system reduced fuel consumption by 22-31% for acceleration conditions and by 12-26% for deceleration conditions. Several studies also reported that human-driven vehicles may benefit environmentally while driving near or behind eco-driving vehicles (e.g., [*34*]).

The potential benefits of eco-driving can reach an upper bound in WZs as they typically induce speed variation, lane change, acceleration and deceleration, and driving aggressiveness. However, eco-driving benefits can be challenged at WZs that experience traffic congestion and have limited MPR. Mensing et al. [*35*] found that eco-driving benefits dropped by 16-54% to preserve safe car following in congested traffic conditions. Rios-Torres et al. [*36*] also found that a limited MPR only reduced fuel consumption at low traffic volume whereas an increase in fuel consumption resulted at high traffic volume. Moreover, Mersky et al. [*37*] reported that without connectivity or predictive features automation alone may generate higher fuel consumption to maintain the safe following distance. In the absence of communication with human-driven vehicles or if CAV anticipatory features become inactive, slow, or unreliable, CAVs will rely on their sensors and may be forced to decelerate or accelerate sharply to maintain safety resulting in more energy consumption. At congested WZs, CAVs will not only be concerned about the sudden stop-and-go manoeuvres of human-driven and unconnected vehicles but also about their uncoordinated lane changes.

### *2.3.2 Platooning (Automation- and Connectivity-Based)*

Platooning reduces aerodynamic drag and allows vehicles to use energy more efficiently and reduce emissions. According to several studies [e.g., *30, 31, 38, 39*], conditions that maximize the environmental benefits of platooning include travelling at high speed, adding more members or vehicles to the platoon, and maintaining shorter gaps. Recalling from *Section 2.1.2* that platooning should be prohibited or at least limited on the closed lanes and the first open lane at WZs, platooning environmental contributions can ideally be achieved at WZs with light to moderate traffic demand and on the far open lanes away from the closure. Full CAV penetration (100% MPR) and efficient platoon lane-change algorithms are needed to preserve platoons on the closed lanes and at congested WZs.

Platooning of trucks is especially beneficial environmentally. Tsugawa et al. [*39*] found through a field test that a platoon of three heavy trucks yielded 16% fuel saving at a 4.7-m gap and this reduced to 9% at a 20-m gap. On multi-lane WZs with a high percentage of trucks, truck platooning can be encouraged by restricting truck operations to specific lanes. If only one lane remains open, rerouting strategies or detours may target only passenger cars, so the trucks remain together on the original path.

### *2.3.3  Eco-Lane (Mainly Connectivity-Based)*

Gathering all eco-vehicles in one lane can elevate the MPR locally and regain the environmental benefits that are only or ideally achievable with high MPRs. This eco-lane can be especially advantageous during the early stages of CAV deployment when low MPRs will be on the roads. As compared to a fully dedicated CAV lane (*discussed in Section 2.1.3*), the criteria to gather vehicles in an eco-lane are environmental-oriented. Eco-vehicles will be instructed to follow recommended speeds and to adhere to maximum acceleration, these parameters should be updated based on the real-time and near-future traffic conditions and can be relayed to vehicles through V2I (RSUs) or through V2V communications. Thus, vehicles must be connected to contribute to the eco-lane but may not be automated. Ahn et al. [*40*] developed and tested an eco-lane system using microscopic traffic simulation, a significant reduction in fuel consumption and emissions was reported. The study also recommended that the system can be pre-emptive, i.e., when congestion becomes imminent, the system is activated and eco-vehicles are instructed to formulate the eco-lane in advance. The eco-lane is better to be placed at WZs on the open side of the driveway away from the closure to reduce the friction with lane changes made by human-driven vehicles. The eco-lane should ideally be activated at medium to high traffic demand levels and when vehicles start approaching the WZ area, i.e., when speed variation becomes high. Setting the spatial and temporal boundaries of the eco-lane in detail at WZs deserves future research.

### *2.3.4  Eco-Routing (Mainly Connectivity-Based)*

The objective of eco-routing is to guide the vehicles to the routes that result in reducing their environmental impacts, i.e., minimizing fuel consumption and emissions. Eco-routing has a great potential at congested WZs even with low MPR. Eco-routing that results in a slim percentage reduction in traffic demand on the congested route can significantly reduce the chance of traffic breakdown and avoid the resulting stops, braking, and acceleration on that route, or at least it can significantly lessen the level of congestion. Kim et al. [*41*] found that fuel consumption and emissions increased by more than 80% when the traffic transferred from free-flow to heavy congested conditions at WZs. Even mitigating the WZ congestion from heavy (5 mph average speed) to medium (25 mph) was very beneficial, fuel consumption and emissions were reduced by 40%.

Guo et al. [*42*] introduced a selective eco-routing strategy whereby vehicles with the largest potential of energy savings or emissions reduction are selected for the alternative routes rather than being randomly selected, i.e., travellers are dynamically ranked according to their location in relation to the alternate routes and only top drivers are selected to achieve a greener equilibrium. This selectivity makes eco-routing very efficient even at very low MPRs. For that case study, when the MPR reached 40% the environmental benefits almost were maximized with 12% reduction in CO emission. At 10% MPR, the reduction in CO emission dropped to only around 7%. Lane change maneuvers at WZs impose additional acceleration and deceleration and can considerably increase fuel consumption and emissions. By adopting a V2I-based eco-routing strategy that selectively targets CAVs on the closed lanes, the number of lane change manoeuvres and the resulting friction can be substantially reduced.

### *2.3.5  Speed Harmonization and Shock Wave Damping (Automation- and Connectivity-Based)*

WZ areas are associated with a large speed variation, thus speed harmonization can be very effective in reducing the energy consumption and emissions when it maintains a moderate average speed and limits the unnecessary acceleration and deceleration [*25*]. Ghiasi et al. [*43*] found that the energy efficiency attributed to speed harmonization increases by adding automation to connectivity (i.e., more CAVs against only connected vehicles). In addition, speed harmonization has also the potential to dampen and absorb shockwaves during congested conditions. Stern et al. [*44*], for example, field-tested the impact of small MPR of automated vehicles on shock wave damping. They used a circular ring road with 20 human-piloted vehicles and a single autonomous vehicle. During shock-wave conditions, the automated vehicle was efficient in reducing the emissions of the whole traffic stream by 15% ($CO_2$) and 73% ($NO_x$); the vehicle kept a smart gap and speed with the leading vehicle and used this gap to avoid sharp braking when the leading vehicle entered a shock wave. This benefitted the whole traffic fleet and absorbed the shockwaves. The study however noted that lane change maneuvers on real highways may restrict the ability of automated vehicles to dampen shockwaves, a subject that merits further investigation.

*2.3.6 Cooperation and Anticipation of CAVs (Mainly Connectivity-Based)*

Lane changes induce significant additional braking and acceleration not only by the lane-changer, but also by the nearby vehicles leading to added fuel consumption and emissions. Several efforts [*e.g., 45-47*] developed and evaluated cooperative lane-change algorithms for CAVs on basic highway sections or at on-ramp merge junctions and reported significant reduction in fuel consumption due to the resulting smoother flow and the shorter lane change time. The cooperative lane change model developed by Choi et al. [*47*] was evaluated by several levels of service and by two scenarios: merging from a slower to a faster lane and vice versa. When cooperating, the following vehicle on the target lane typically reduces speed to create a safe merging gap to the lane-changer. The cooperation reduced emissions significantly; however, the reduction was larger for the scenario of lane changing from the slower to the faster lane (i.e., as the case of WZ lane changing) and was also more prominent at lower levels of service (i.e., larger traffic demand). The largest observed reduction in emission was 11.8% for slow-to-fast lane change at LOS E. Li et al. [*48*] found that early advisory messages to change lane at WZs were able to reduce the time and the distance needed to change lane by 28% and 27% respectively. This study relied upon communicating with human drivers in a driving simulator, the findings therefore clearly show a great potential to save energy when anticipating and preparing for the lance change in advance of the WZ taper and even without automation. According to the existing literature, WZs that accommodate cooperative lane change can gain significant environmental benefits. However, a major challenge to this is the MPR, most of the existing studies that promoted cooperative lane-change assumed full CAV penetration, which calls for further efforts that investigate various penetration levels.

**2.4 Demand of Automation and Connectivity at WZs**

Figure 7 outlines the demand level of automation and connectivity in the WZ vicinity, which is a high-level analysis inferred from the previously discussed concepts. Generally, the need for automation increases as vehicles move downstream whereas the need for connectivity is highest upstream. In the WZ approach area and further upstream, vehicle functions are mainly connectivity-based (e.g., rerouting, early merging, receiving WZ and end-of-queue warnings, forming ideal flow distribution, forming eco-lane or CAV-lane). The congestion level is usually lower upstream with no queuing, vehicles do not need advanced automation there because driving is less stressful. At both the WZ queuing and activity areas, both connectivity (hazard detection, collision warnings, platooning, cooperation) and automation (tight coupling, stop-and-go navigation, eliminating drivers' errors and imperfections) are needed. Finally, downstream the WZ area, automation is needed to discharge the queues effectively by controlling the acceleration and the gaps, the demand for connectivity decreases but it is moderately needed to enhance vehicles clustering and their synchronized acceleration.

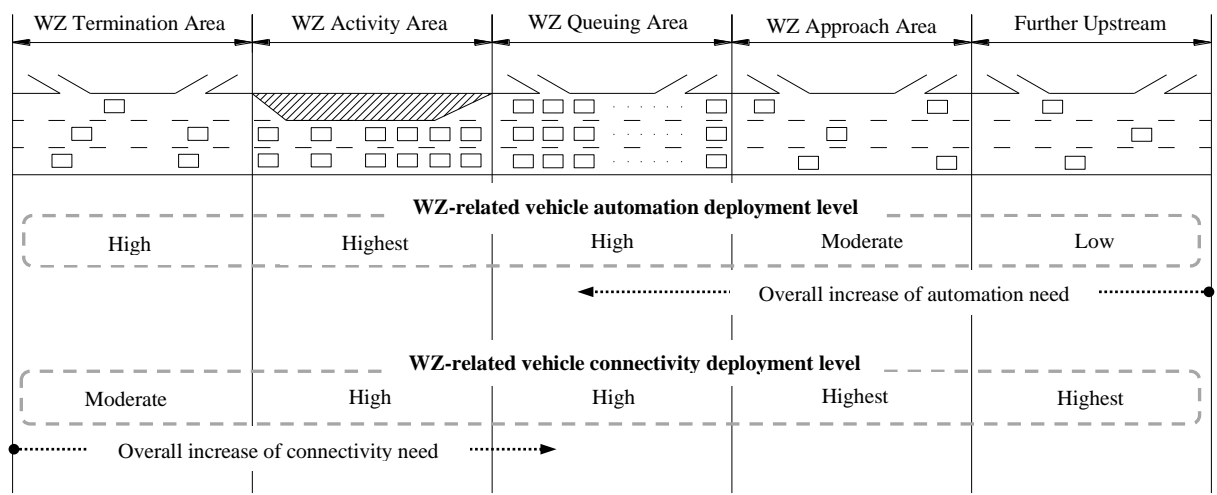

**Figure 7. Level of deployment of automation and connectivity at WZs.**

**2.5 CAV Impact on WZ Traffic Performance: Existing Literature**

Table 1 summarizes the main efforts found in the literature that assessed the impact of CAV systems on WZ traffic. This table only considers studies that assessed traffic flow performance, studies that used specific technologies for specific purposes at WZs were discussed elsewhere (i.e., *Section 3*). The studies so far focused on safety and mobility aspects, no studies were found that particularly examined environmental aspects or driver's behaviour related to CAV operations.

At this stage when only few efforts exist, conclusive quantitative results cannot be drawn reliably but some indicative or general lessons can be learned. Overall, the reported size of CAV benefits at WZs appears to be large, but this was critically sensitive to the MPR and the level of traffic flow at WZ. Zou et al. [*49*], for example, found that CAV MPR of 34.1%, 62.25%, and 100% achieved average travel time reductions of 25%, 50%, and 90% respectively at a congested 2-to-1 WZ configuration. Abdulsattar et al. [*50*] tested the impact of connected vehicles at several traffic demand levels. For example, for a demand level of 3000 vph and MPR of 75% of connected vehicles, the mean travel time was reduced by 40% and capacity increased by 65% at 2-to-1 WZ. Only at high traffic demand levels the benefits of connected vehicles were significant. In addition, for each traffic flow level, there was a critical range of MPR where the benefits of connected vehicles became maximized and after which there would be no more feasible benefits by increasing the MPR. Genders et al. [*53*] evaluated the safety benefits resulting from rerouting connected vehicles away from WZs. The time-to-collision was used as a safety surrogate measure. Moderate MPRs (<40%) enhanced the network safety because the improved driving behaviour overweigh travelling additional trip distances; however, the network safety degraded for large MPRs (>40%) because substantially longer trip distances were added to the network and this increased the exposure to safety hazards and made the enhanced driving behaviour less effective. Abdulsattar et al. [*54*] examined the impact of connectivity on reducing rear-end collisions and indicated that the first entry-level of 10% MPR resulted in significant safety improvement. For example, at medium and high traffic flow rates, the first 10% MPR decreased the critical time-to-collision by 50% at 2-to-1 WZ, i.e., the probability of rear-end collision was reduced by 50%. Beyond the first 10% MPR, higher traffic flow rates required higher MPRs to show further safety improvements. More studies and findings can be learned from Table 1.

The cited studies overall indicated promisingly positive results, e.g., capacity enhancement, travel time and delay reduction, smoother and safer merges, harmonized speed, enhanced driver's awareness, and reduced chance of rear-end collision. The existing literature however only explored a limited number of CAV concepts. More efforts are needed and recommended to explore all the mobility, safety, and environmental concepts discussed in the previous sections. Combining several CAV concepts at one WZ also merits investigation. Nevertheless, attention must be paid to the driving behaviour used or assumed in the analysis. The cited mobility studies, for example, were generally interested in assertive driving behaviour and shorter following gaps whereas safety studies tended to assume conservative following behaviour and longer gaps. This indicates that CAV benefits are very sensitive to the assumed driving behaviour calling for more efforts that attempt to understand driver's behaviour of CAVs and conventional vehicles at WZs. Most of the cited studies relied on simulation tools in the absence of considerable CAV presence in the field currently. Studies should however consider field testing as well in the short-term, e.g., by using testbeds and naturalistic driving experiments.

**Table 1. Literature Review Summary for CAV Impact on WZ Traffic Performance**

| Authors | Main Theme | Study Scope | Analysis Tool | Main Findings |
|---|---|---|---|---|
| **Zou et al. [*49*]** | Mobility | Explored how CAVs generally would impact several WZ traffic operational measures (travel time, emissions, speed harmony). | Cooperative Cellular Automata Model (CCAM) simulation | CAV deployment reduced travel time and emissions and increased speed harmony. Larger MPRs enhanced these benefits and reduced the stochasticity in the measured variables. |
| **Abdulsattar et al. [*50*]** | Mobility | Explored the impact of vehicle connectivity (V2I and V2V) on WZ traffic speed, capacity, and travel time and its variability. | Agent-based modelling and simulation (ABMS) | The benefits of connected vehicles (larger capacity, higher speed, and lower mean travel time) become noticeable and significant only at high traffic flow rates. |
| **Ren et al. [*15*]** | Mobility | Proposed and evaluated a collaborative merge control technique at WZs. | Vissim microscopic simulation software | The cooperative merge control technique proved to outperform other traditional merge control systems. For mixed traffic, the applicability of the technique needs further investigation. |
| **Weng et al. [*51*]** | Mobility and Safety | Proposed a merging assistance system at WZs that guides safe merging. | Used vehicle trajectory data and non-paramtric method called "CART" | The most influencing factors on merging behaviour were the merging time elapsed, remaining distance to WZ, speed, and the time-to-collision between the merging vehicle and neighbouring vehicles. |
| **Ramezani et al. [*52*]** | Mobility and Safety | Developed and evaluated a connected-vehicles-based speed harmonizer methodology at WZs. | LOQO solver for large-scale nonlinear optimization | The delay was reduced by 13% when the MPR was 80% or higher. During 40-minute operation period and 100% MPR, the congestion period was reduced by 26.4%. |
| **Genders et al. [*53*]** | Safety | Evaluated the impact of providing connected vehicles with warning messages of the WZ presence on the network time-to-collision (TTC). | Paramics microscopic simulation software | Early dynamic rerouting reduced driver's agressiveness but this also increased average trip distance which in turn enlarged the exposure to safety hazards especially with high MPR. |
| **Abdulsattar et al. [*54*]** | Safety | Evaluated the impact of deploying vehicle connectivity on the probability of rear-end crashes at WZs. | Agent-based modelling and simulation (ABMS) | The first entry-level of 10% MPR resulted in substantial safety improvement. Beyond the first 10% MPR, higher traffic flow rates required higher MPRs to show further safety improvements. |
| **Qiao et al. [*55*]** | Safety | Evaluated smartphone-based rear-end collision warning messages at WZs. | Driving simulator environment | Voice and sound warnings were effective at reducing speeds and prolonging headways. Visual warning deteriorated driving behaviour and generated visual distraction. |

### 2.6 Conflict between Mobility, Safety, and Environment Objectives

Efficiency of traffic flow performance at WZs should be assessed by tracking several objectives including mobility, safety, environment, and driver's comfort. In many cases, these objectives conflict and do not accord. For example, mobility requires narrowing the headways to optimize capacity, but safety calls for preserving long headways to reduce the chance of rear-end crashes. Several rerouting studies found that the travel-time-equilibrium does not necessarily coincide with the green-equilibrium [*42, 56-58*], they reported several scenarios where reducing travel time can increase emissions and fuel consumption and vice versa. Wadud et al. [*31*] also warned that eco-driving may lead to an adverse mobility impact because driving with moderate speed and gentle acceleration may decrease capacity and cause congestion. Safe car-following needs long inter-vehicle headways and allowing the use of sharp acceleration or deceleration to respond to sudden conflicts promptly; eco-driving however calls for shorter headways (to reduce aerodynamic drag) and gentle braking and throttle control. Several studies [*37, 59*] reported that cautious and conservative car-following algorithms increase fuel consumption and emissions. Mersky et al. [*37*] and Makridis et al. [*60*] also found that automation alone (without connectivity) increases fuel consumption due to losing the anticipatory features in the car-following algorithms.

The response to the potential conflicts discussed ahead is to develop multi-objective car-following and routing algorithms that consider mobility, safety, environment, and driver's comfort. Wu et al. [*61*], for example, used a car-following algorithm that is based on fuel efficiency, but the system switches to a driver safety mode when the headway becomes less than a certain threshold. Li et al. [*62*] developed an adaptive cruise control system that considers fuel economy, safety, and driver's comfort. Djavadian et al. [*63*] developed a dynamic multi-objective eco-routing strategy that explicitly incorporates reducing travel time and environmental pollutants. Although these studies had some limitations, i.e., they assumed full MPR and they used non-exclusive multi-objective functions, they indicated promising results in terms of mitigating the conflicts between several objectives and bringing the traffic flow to a universal optimum point. Studies are certainly needed to develop multi-objective car-following algorithms at WZs while considering the peculiarities of these locations (geometric complexity, lane change, queuing, congestion, detours, etc). In the near term with low MPRs, safety should be given more weight against other objectives. As CAV driving becomes more reliable, the safety weight can be relaxed in favour of the other objectives.

**2.7 The Role of Driver's Behaviour**

Driver behaviour at WZs should also be assessed carefully to ensure that driver's comfort is met and that it does not conflict with other traffic flow performance objectives. A major behavioural challenge at WZs is that CAV drivers must be assured of CAV systems and their ability to operate reliably and safely within the WZ vicinity albeit its complexity and the needed lane changes. Robertson et al. [*64*] indicated that at least a significant proportion of CAV drivers may prefer to switch off the automated features. If drivers remain cautious or doubtful, many may switch the automation off when approaching the WZ and this can result in lowering the effective MPR; consequently, CAV benefits will be universally degraded. Another behavioural challenge that may impact many mobility and environmental concepts is that many CAV drivers may insist on having cautious and long headways while traversing WZ areas especially when travelling nearby human-driven or unconnected vehicles.

If CAVs are not fully automated (e.g., levels 2 or 3), i.e., a common expected scenario during the early penetration phase, drivers may also be concerned about when and how many times they will need to retake the control from the vehicle when traversing the WZ area. Recent research findings [*65, 66*] revealed that drivers may perform poorly and show low situational awareness when retaking the control from the vehicle. Merat et al. [*65*] found that drivers took 10-15 s to resume control after disengaging the automation and 35-40 s to stabilize their lateral control of the vehicle; this can be very critical at a lane-change zone. If drivers learn by experience that automation cannot be continuously maintained along the WZ vicinity, due to limited technology or low MPR, some may consider travelling under monotonous and intermittent calls to resume vehicle control more stressful and less safe as compared to driving manually the whole WZ segment. This can be a special case at long WZs with varying geometry.

On the other hand, the literature provides some evidence that drivers are more in favour of switching to automation during congestion, but under high or full automation levels (i.e., level 4 or 5). Payre et al. [*67*] found that traffic congestion motivates drivers to switch to full automation to avoid boredom and stressful driving. WZs are well known to cause congestion, delay, and stress. With full automation and high MPRs, drivers are expected to switch control to the vehicles so they can be relieved from stress and also exempted from liability. Drivers still appear to keep additional attention to the roadway under congested conditions as compared to non-congested conditions [*68*].

Drivers' behaviour of conventional vehicles in a mixed traffic WZ is not less important. If CAVs cooperate to form long and frequent clusters on the open lanes, nearby conventional drivers will be less prioritized and may start feeling stressed and be tempted to perform aggressive manoeuvres. This problem may exacerbate if the clusters are mainly truck-platoons which occupy larger space and block longer visibility. The behaviour of conventional drivers raises many questions. Can conventional drivers discern the type of nearby vehicles, i.e., connected, automated, or conventional? How conventional drivers will behave if they learn that CAVs yield more or less courteously to their aggressive merges as compared to conventional vehicles? Would they cooperate with CAVs?

Research efforts that can evaluate the behaviour of CAV systems and drivers of CAVs and human-driven vehicles at WZs are needed. The available literature lacks such efforts which are very important to calibrate car-following and lane-changing algorithms and to preserve driver's comfort while traversing WZs.

## 3. TECHNOLOGIES AND INFRASTRUCTURE CONSIDERATIONS

The previous sections discussed the potential of CAVs to improve several traffic flow measures. The interaction between CAVs and WZs is however not limited to traffic flow and behaviour, it also brings many technological and infrastructure implications, challenges, and opportunities. This part of the paper reviews the technological and infrastructure considerations related to CAV deployment at WZs. CAV technologies represent a complex, interdisciplinary, and rapidly evolving area of research; it is therefore a very challenging task to review all related subjects and predict their future in one paper. This section attempts however to discuss the state-of-the-art of the most important and pertinent topics to WZs including detection of WZ features, smart traffic control devices, communication technologies and types, real-time tracking of WZ actors, wearable technologies, guidance of pedestrians and cyclists, WZ data systemization, and asset monitoring.

### 3.1 Detection of WZ Setting and Devices

An important navigation challenge for CAVs at WZs is the sudden change in lane configuration and the resulting complex geometry. CAVs cannot solely rely on high-definition digital maps, these may not be updated promptly. This is especially a problem for short-term WZs that may last only a few days or even hours. CAV detection of traffic control devices (TCDs) at WZs is therefore indispensable. WZ TCDs differ from the devices used on standard roadway sections because they usually include signs with more complex contents, temporary channelizing devices (cones, barrels, markings), flaggers, and others. CAVs need to perceive these devices reliably and promptly so they can traverse the WZ efficiently and safely. At high automation levels, vehicles should autonomously recognize and react to the TCDs. Even when drivers are in control of the vehicle, vehicle-based detection of TCDs can assist them by providing prompt in-vehicle visual or voice notification of the device. The following sections provide an overview of several concepts that facilitate the interaction between CAVs and several types of TCDs at WZs.

#### 3.1.1 Camera-Based Detection

Camera-based detection of TCDs is a well-established research area that uses sophisticated and computerized vision-based techniques. Under these systems, the vehicles use real-time videos of the roadway ahead. The video images are then analysed using machine learning techniques (e.g., neural network [*69, 70*], deep learning [*71*], support vector machine [*72*], and random forest [*73*]) that classify and recognize the device based on an inventory of predefined and standardized devices. Wang et al. [*74*] developed a camera-based system to detect temporary traffic cones, the system accurately recognized and detected the cones with success rates of 100%, 100%, and 85% for the blue, yellow, and red cones, respectively. The author noted that the recognition of the red cones was possibly impacted by the experiment background colour. The same system was also able to maintain 90% accuracy when sensing the cone distance. Lee et al. [*75*] tested a Kernel camera-based sign detection system at WZs, the system reduced false detection rates as compared to other vision systems. Some studies [*e.g., 76*] showed that lane markings can also be recognized by camera systems if they were standardized (i.e., with specific shape, dimension, and color). According to the literature [*e.g., 75, 77, 78*], challenges of the camera-based detection include: (i) the lack of a standard database of TCDs that can be generalized everywhere, (ii) the system performance may change by different road, weather, and illumination conditions, and (iii) the likelihood that the system may detect a device that does not belong to the vehicle route (e.g., signs on intersecting roads). Furthermore, camera-based detection systems usually have a limited database capacity and WZ TCDs may not be given priority since they treat temporary conditions.

#### 3.1.2 LiDAR-Based Detection

Light detection and ranging (LiDAR) is a remote sensing technology that illuminates target objects with laser light and the reflected light is then detected and analyzed by sensors. The sensors can measure distances and create coloured 3D images which enable LiDAR-equipped vehicles to locate obstacles and recognize TCDs. LiDAR detection is also a computerized vision-based method, but it infers the images using the reflective intensity data and not video recordings. Several recent efforts indicated promising results for using LiDAR to detect TCDs [*e.g., 79-82*]. These studies also emphasized that the reliance of LiDAR on the device retro-reflectivity features provides more detection robustness as compared with camera-based detection methods which can be impacted by light conditions, image quality, and occlusions. This allows to detect more types of TCDs and to manage a larger inventory of traffic signs. LiDAR detection also provides robust 3D information on the surrounding setting. Furthermore, LiDAR-based detection has been reported as a promising technique to detect and track lane markings [*e.g., 83-85*], LiDAR appeared less sensitive to shadow, sunlight, or night as compared to camera-based detection. The LiDAR cloud images do not necessarily rely solely upon the painting's retroreflectivity. However, the retroreflectivity of the marking paint can make its laser reflectivity value larger than the surrounding and hence more discriminable. However, one challenge to LiDAR is that laser reflectivity can be impacted by the device or sign pose. Some researchers recommended a hybrid system that relies on the fusion of complementary data from two sensors, i.e., a camera and LiDAR, in order to further improve the robustness of the detection algorithms [*e.g., 86*].

### *3.1.3 Communication-Based Detection*

A V2X communication can be established between a vehicle and a device equipped with a tag that transmits a message defining the device identity. Garcia-Garrido et al. [*78*] proposed a radio frequency V2X communication system that uses wireless sensors installed on the traffic signpost; the sensors transmit a message to an on-board unit (OBU) including information about the sign type, its position, and the road name. The sign was able to communicate to all equipped-vehicles within a 500 m coverage, the study also indicated the possibility to widen the coverage. Qiao et al. [*87*] and Liao et al. [*88*] similarly developed and successfully tested a V2X sign detection system at WZs using radio frequency and Bluetooth respectively. Connected cones, barrels, and panels are also in use today (*Section 3.2.1*) and can easily communicate with CAVs. Communication-based detection of lane marking is also possible by using radio frequency identification chips either embedded in the pavement or placed as lane dividers [*6*], such smart lane marking techniques can also be deployed to locate other nearby CAVs making those functions that rely on accurate positioning of other vehicles more efficient.

The communication-based detection of TCDs appears more reliable for CAV operations at WZs, it resolves the limitations of the machine-vision detection systems; i.e., it discards irrelevant devices, it eliminates the chance for false detection, and it customizes the device identification (message content and delivery time) according to each site characteristics. However, cost-effectiveness of connected TCDs is yet to be assessed.

### **3.2 Smart Traffic Control Devices**

### *3.2.1 Connected Traffic Control Devices (Connected TCDs)*

Connected TCDs refer to devices that are supplied with sensors and connected wirelessly. The market of these devices is growing [*89*] and it includes connected cones, barrels, panels, and arrow boards. The connectivity of these devices is deployed today in various applications [*e.g., 89-92*]; the sensors can provide real-time data to validate WZ geometry, detect errant vehicles, send a request to remove the debris and reinstall cones when a vehicle hits the WZ taper, send a voice or visual alarms to the workers on site if a hazard is detected, and can also detect pedestrians trespassing the WZ area. Nevertheless, the functionality of these devices can be easily extended by establishing a communication with approaching CAVs which can subsequently easily discern the devices and identify the WZ boundary [*90*]. Transitioning from traditional to connected TCDs may therefore become more warranted with the forthcoming penetration of CAVs, and their market is expected to continue to grow.

### *3.2.2 Robotic Traffic Control Devices (RTCDs)*

RTCDs are intelligent devices that can move autonomously while being connected to a central robot base. RTCDs mainly include barrels, cones, barricades, and sign bases. Changing the positions of the RTCDs dynamically, flexibly, and remotely based on the work activities can reduce the size and duration of WZs and reduce workers' exposure to traffic. RTCDs must move reliably, they may otherwise generate hazards. Shen et al. [*93*] developed and tested robotic safety barrels, which move autonomously at WZs using a central station which localizes the barrels and communicates the planned path for each. The system was tested in realistic highway environments and the maximum final positioning error for all robots was 11 cm exceeding human accuracy of barrel placement. The system was intended to operate with human-driven vehicles. However, CAVs can easily communicate with RTCDs when approaching the WZ area through V2I or V2X channels.

*3.2.3 Automated Flaggers*

Automated flagger assistance devices (AFADs) have been already applied in some states and countries at WZs [*94-96*] allowing to operate an automated device (e.g., Stop-Slow signs, Red-Amber lights) from a communication base away from the roadway. The deployment of these devices initially aimed at enhancing safety by reducing human exposure to traffic and reducing labour cost if applied for long durations. Some challenges of AFAD deployment were noticed among researchers, e.g., drivers had difficulty to understand the system and both the approaching speed and the deceleration rate increased raising the risk of rear-end collisions [*94-96*]. However, by establishing a V2I communication between CAVs and AFADs, the deployment of AFADs at WZs can be greatly revitalized. CAVs cannot communicate easily with human flaggers and therefore they need AFADs especially under high automation levels. Moreover, CAVs would receive standard and prompt messages that allow for easier understanding of the system and smoother deceleration.

**3.3 Communication Technologies at WZs**

This section discusses the communication technological aspects between WZs and CAVs. First, the main potential communication technologies at WZs are introduced. Communication types and needs are presented afterwards. Finally, the available literature that investigated CAV communication at WZs is summarized.

*3.3.1 Communication Technologies*

Two important parameters of communication technologies for transportation applications are latency (time between when the information becomes available for broadcast and when it is received) and range (distance between two communicating units needed to have an efficient communication). Low latency and long range are desired, but unfortunately they cannot coexist easily, technologies that offer low latency suffer from narrow range and vice versa. Practitioners must choose the communication technology based on the needed applications.

*3.3.1.1 Dedicated Short-Range Communication (DSRC)*

Many studies considered DSRC as the most promising communication technology for CAV applications in the near term [*97-101*], this is valid for V2I, V2V, and hybrid communications. DSRC has many advantages that fit transportation-oriented applications well, including high reliability, low latency, interoperability, and security. The most fitting applications to DSRC are safety and crash warnings. The literature agreed that crash warning applications require latency as low as 100 ms [*e.g., 102*], this makes most non-DSRC communication technologies inefficient for WZ crash warnings. Vehicle clustering and tight coupling at WZ also need low latency and DSRC capabilities. However, one main disadvantage of DSRC is the limited range and scalability.

*3.3.1.2 Cellular Communication*

Capabilities of early cellular technologies are opposite to those of DSRC, i.e., they offer wider range and larger packet data but with higher latency. Xu et al. [*103*] compared 4G-LTE cellular technology with DSRC and found that DSRC clearly outperformed 4G-LTE for safety application; however, due to its wider availability and higher throughput, the 4G-LTE was recommended for applications that require information transmission and file download. However, the rapid evolution of the cell-phone consumer market has made this technology more deployable. Recent studies indicated that the newly introduced 5G-based communication technologies have small latency and can be deployed efficiently for CAV communications and applications [*e.g., 104-108*]. The arrival of 5G-cellular communication has enhanced the cellular vehicle-to-everything communication concept (C-V2X) which comprises communication types beyond V2V and V2I, e.g., V2P (vehicle-to-pedestrian), V2D (vehicle-to-device), and V2N (vehicle-to-network). The 5G C-V2X appears to be a possible competitor to DSRC for time-critical applications especially if it succeeds in combining low latency with wide coverage.

*3.3.1.3 Satellite Communication*

Global Positioning System (GPS) communication technologies offer massive coverage (hundreds of kms) but with very high latency (can reach 10-20 seconds). However, they can still be deployed in some mobility applications such as WZ presence warning, weather alerts, and alternate route advisory.

Dehman and Farooq 20

Constellations of low earth orbit (LOE) satellites (e.g., SpaceX and One-Web) are growing rapidly and they may revolutionize satellite-based communications. These mega-constellations have a large number of satellites (hundreds or thousands) that are close to earth allowing for lower latency, precise positioning that may outperform the GPS, and broadband internet connectivity that can support various in-vehicle functions [*109-112*].

### *3.3.2 Communication Types and Needs*

Figure 8 illustrates the main communication types at WZs including V2I, V2V, and V2X. Table 2 also outlines the essential communication needs for all the aforementioned CAV functions and concepts pertaining to mobility, safety, environment, and TCDs. This table was based on the essential functions of each application and the available literature. Nevertheless, exceptions may be observed for specific systems and the table may be further fine-tuned when more advanced technologies and more research findings become available.

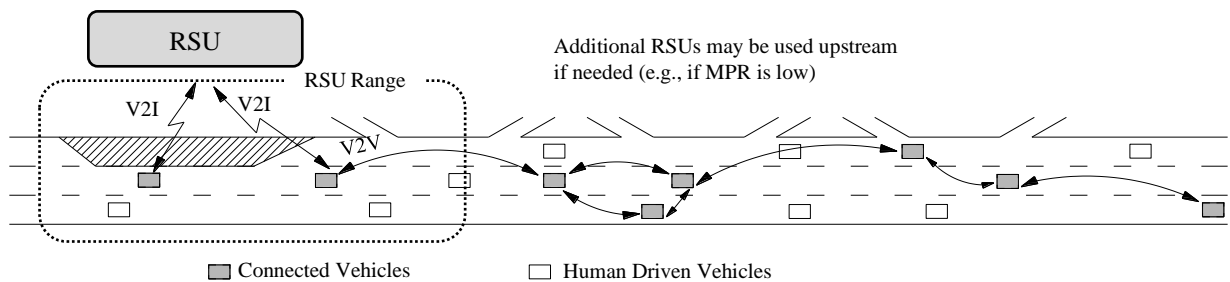

(a) Basic communication types (V2I and V2V)

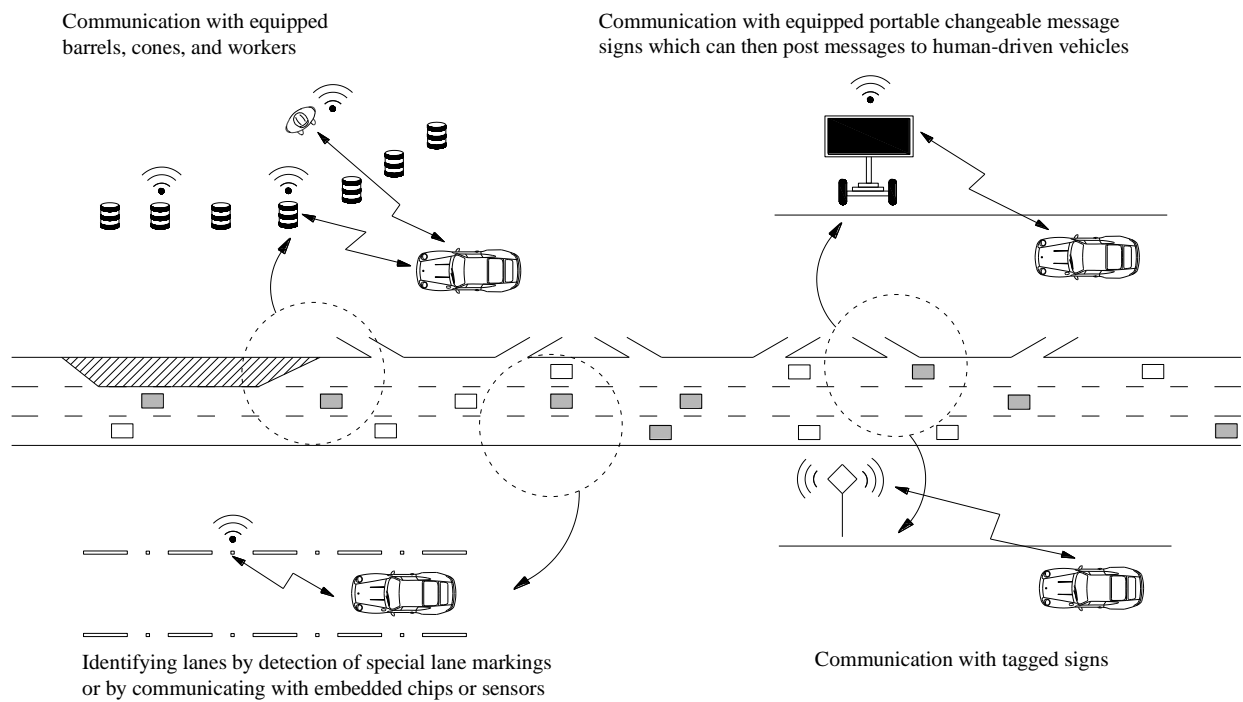

(b) Additional V2X communication types

**Figure 8. Different CAV communication types at WZs.**

**Table 2. Communication Needs for Some CAV Functions at WZs**

| Function | Essential Communication Needs | | |
| --- | --- | --- | --- |
| | **Type** | **Desired Features** | **Example Technology** |
| Tight coupling | V2V | Low latency, high reliability and accuracy | DSRC, 5G Cellular* |
| CAV clustering | V2V | Low latency, high reliability and accuracy | DSRC, 5G Cellular* |
| Dedicated CAV-lane & eco-lane | V2I | Wide range | Cellular, Satellite, and Scalable-DSRC |
| Synchronized acceleration | V2V and can be enhanced with Hybrid V2I+V2V | Low latency, high reliability and accuracy | DSRC, 5G Cellular* |
| Cooperative lane change | V2V | Low latency, high reliability and accuracy | DSRC, 5G Cellular* |
| Smart lane flow distribution | V2I | Wide range | Cellular, Satellite, and Scalable-DSRC |
| Early rerouting & eco-routing | V2I | Wide range | Cellular, Satellite, and Scalable-DSRC |
| Vehicle automation | V2V | Low latency, high reliability and accuracy | DSRC, 5G Cellular* |
| Speed harmonization | V2I, V2V, or Hybrid V2I+V2V | Wide range | Cellular, Satellite, and Scalable-DSRC |
| Work zone presence warning | V2I | Wide range | Cellular, Satellite, and Scalable-DSRC |
| Rear-end warning | V2V | Low latency, high reliability and accuracy | DSRC, 5G Cellular* |
| Sideswipe warning | V2V | Low latency, high reliability and accuracy | DSRC, 5G Cellular* |
| Work-activity area crash warning | V2X | Low latency, high reliability and accuracy | DSRC, 5G Cellular* |
| Vision-based detection of TCDs (camera-based, LiDAR-based, or hybrid) | Vehicles are equipped with machine-vision detection systems, communication is not needed | | |
| Communication-based detection of TCDs | V2X | Low latency, high reliability and accuracy | DSRC, 5G Cellular* |
| Automated flaggers | V2I | Low latency, high reliability and accuracy | DSRC, 5G Cellular* |

(*) needs further confirming research and testing

*3.3.3   CAV Communication at WZs: Existing Literature*

Table 3 summarizes the main existing literature efforts that designed or tested CAV communication technologies at WZs. The Bluetooth and Ultra-Wide Band technologies provided limited range [*88, 116*] making them only applicable for short-range applications like sign detection. The cellular internet communication allowed for a network-wide coverage but with limited latency [*115*]; however, as stated previously the recent 5G cellular technology promises for low latency and deserves further testing at WZ sites. The one well researched technology at WZs that provided a trade-off between coverage and latency is the DSRC as explored by Maitipe et al. [*100, 101*], Zaman et al. [*113*], and Ibrahim et al. [*114*]. In these studies, approaching drivers were provided with information regarding the expected travel time and congestion location (queue length upstream the WZ). It is useful to summarize below the lessons learned from these studies which enhance the transferability and capability of DSRC systems at WZs.

*Scalability of DSRC System*

The V2I-DSRC system range can be increased by using several RSUs systematically spaced upstream the WZ. For V2V-supported DSRC systems, the communicated messages can reach any upstream distance as long as there are equipped-vehicles that continue to relay the message (i.e., sufficient MPR). To account for low MPR, vehicles in the opposite direction may also be incorporated in the rebroadcast channel [*113*].

*Communication Load*

In dense traffic, communication may create a broadcast storm if each vehicle receiving a message is relaying it forward. This can be important for WZ areas where dense traffic is common. A "Selective Relay" protocol was used to ensure that only one vehicle, selected in the farthest band distance from the last-originating vehicle, will relay the message [*100*]. The route can also be divided into concatenated rectangles and subrectangles, and only the farthest vehicle from the beginning of each subrectangular relays the message [*113*].

*Communication Region*

Communication should only be maintained on the WZ roadway, messages should not be exchanged with vehicles on parallel or intersecting roads. Maitipe et al. [*100*], established protocols to define an angular region around the WZ roadway and to check the road horizontal curvature, vehicles travelling outside the region were instructed to drop any received message immediately. Zaman et al. [*113*] used geographically-defined concatenated rectangles to predefine the hopping routes, the length of these rectangles decreases when the highway curvature becomes sharp and increases on tangent stretches. If an equipped-vehicle receives a message outside the rectangles, i.e., on parallel or intersecting roads, then it will ignore the message. If desired, upstream on-ramps may be provided with similar predefined rectangles so they become integrated.

*Communication Directionality*

The messages should be propagated in one direction to avoid congestion and confusion. For messages originating from the RSU and propagated upstream, e.g., travel time estimation and end-of-queue warning, the receiving OBU only relays the message if it is farther from the RSU. Conversely, for messages originated from approaching vehicles and destined for the RSU, e.g., vehicle's location and speed, the receiving OBU only relays the message if it is closer to the RSU.

*Required Market Penetration Rate*

The efficiency of WZ communication systems strictly relies on having sufficient MPR. The RSU relies on passing equipped-vehicles to measure real-time information (e.g., travel time and end-of-queue location), it will therefore have to use unrefreshed and obsolete information if no new equipped-vehicles pass by. Additionally, the V2V back-and-forth message propagation will be halfway interrupted if no equipped-vehicle is found to receive and continue relaying the message. Ibrahim et al. [*114*] proposed a model, along with user-friendly charts, that estimates the MPR needed for the DSRC-based V2I and V2V systems at WZs. The required MPR varies according to the desired information update frequency, DSRC range, and traffic flow rate. A higher MPR is needed if the disseminated information is updated more frequently and if DSRC range and traffic flow are low. As an example, an MPR of 20% was needed for a 10-minute information update frequency, 250-meter average OBU range, and traffic flow of 1300 to 1800 vph.

Table 3. Examples of Field-Tested CAV Communication Technologies at WZs

| Authors | Communication | | System Range | System Function |
|---|---|---|---|---|
| | Type | Technology | | |
| **Maitipe et al. [*100*]** | V2I | Dedicated Short-Range Communication (DSRC) | The RSU range extended around 1 km (0.75 km upstream and 0.25 km downstream). The RSU was placed immediately before the transition area. | Informing approaching vehicles of the estimated travel time and end-of-queue location |
| **Maitipe et al. [*101*]** | V2I and V2V | Dedicated Short-Range Communication (DSRC) | The system can monitor congestion length of up to a few kilometers and the message broadcast coverage can reach a few tens of kilometers. The range is strictly limited by the V2V communication and available MPR. | Informing approaching vehicles of the estimated travel time and end-of-queue location |
| **Zaman et al. [*113*]** | V2I and V2V | Dedicated Short-Range Communication (DSRC) | The ranges of the RSU and the OBU were 500 m and 250 m respectively. The range of the message propagation can be extended upstream as long as equipped vehicles are able to continue relaying the message. | Informing approaching vehicles of the estimated travel time and end-of-queue location |
| **Ibrahim et al. [*114*]** | V2I and V2V | Dedicated Short-Range Communication (DSRC) | The system deploys DSRC-equipped portable changeable message signs (PCMSs) that receive and display the broadcasted information benefitting unequipped vehicles as well. The study also provided charts to estimate the required MPR for continuous and uninterrupted communication. | Informing approaching vehicles of the estimated travel time and end-of-queue location |
| **Qiao. et al. [*87*]** | V2X | Radio Frequency Identification (RFID) | The sign warning range was actuated based on the stopping sight distance. Two distances were used ahead of the signs, i.e., 250 and 305 ft upstream. The study did not discuss if a longer range is possible. | Detection of tagged traffic signs |
| **Azadi et al. [*115*]** | V2I | Cellular Internet Data | The range is only limited by the cellular internet data access. The application does not work in dead zones or where the coverage is weak. | WZ data sharing and general smartphone applications (e.g., route planning, travel information, etc) |
| **Liao et al. [*88*]** | V2X | Bluetooth Low Energy (BLE) | Bluetooth tags were detected 125 m and 150 m upstream when travelling at 70 mph and 55 mph respectively. | Detection of tagged traffic signs |
| **Mollenhauer et al. [*116*]** | V2I | Ultra-Wide Band (UWB) | The range of the UWB system was around 80 m. Due to the limited range, the authors did not recommend UWB systems for WZ open settings. | Detection of tagged workers, equipment, and traffic control devices. |

### 3.4 Other Assisting Technologies at Connected WZs

*3.4.1    Real-Time Tracking of WZ Actors*

Object detection and tracking are at the core of automated ground and air vehicle operations, the success of which is driven by the recent progress in the sensing technologies, deep learning algorithms, and cloud computing. In the context of work zones, such technologies have various safety, traffic management, and enforcement applications. Malveaux et al. [*117*] and Kim et al. [*118*] are some recent examples where video sensors mounted on multi-rotor and fixed-wing unmanned aerial vehicles (UAV) together with deep learning and cloud computing were used to monitor the traffic and construction activities in a work zone. Malveaux et al. [*117*] used the system to provide real-time information to the connected vehicles, while Kim et al. [*118*] analyzed safety related potential conflicts between construction equipment, workers, and vehicles.

The UAV-supported real-time tracking of vehicles can also find beneficial applications at closures on two-lane two-way highways. Usually, human flaggers regulate and alternate the two-way traffic on the single open lane, but flaggers have limited visibility especially on curves and cannot optimize operations. If a UAV tracks traffic and queuing in the two directions, it can be connected to an automated flagger or a traffic signal so traffic operations can be accurately prioritized and optimized. This can be a feasible system at short-term and mobile work zones where installing traffic detectors would be a costly and impractical option.

Mollenhauer et al. [*116*] described and field-tested a WZ hazard detection system whereby all WZ actors (i.e., workers, vehicles, equipment, and cones) where tagged and connected. The system was trained, using machine-learning techniques, to recognize and classify the activities and movements of the workers (i.e., jackhammering, walking, rolling, guiding, or random). The trajectories of connected vehicles and equipment were also tracked by the system that accordingly sends warnings to vehicles or workers when a hazard becomes imminent. Hazard detection algorithm was set to be sensitive to the type of undergoing worker's activity, worker's proximity to the WZ border, prediction of actors' movements, highway geometry (curved or straight), WZ barrier type, blind spots, and the vehicle stopping distance (i.e., speed, pavement friction, and reaction time). The system further considers sending a specific escape direction to workers when connected vehicles are only instructed to brake. The demonstration experiments successfully verified real-time data acquisition and the transmission of safety messages. Tagged objects were positioned within an accuracy of approximately 10 cm; however, the system range was limited to 80 m.

While the trained deep learning models are highly accurate, precise, and robust in prediction, key challenges include the requirement of a large amount of labelled data for training and testing the model, high computational power required for training and finding the optimal hyperparameters, and programming skills required to implement such models. In terms of cloud computing, of-the-shelf commercial services could be used that do not require investment in expensive computing server hardware.

*3.4.2    Wearable Sensors and Technologies*

Wearable sensors and technologies are already in use in the construction industry and they include, for example, smart vests, watches, wrist bands, helmets, and shoes [*e.g., 119*]. WZ workers can wear such devices to keep their position and activity tracked, while working in the activity area. Using V2I (RSU) or V2X (direct communication), the workers can also be alarmed when becoming too close to hazards (e.g., errant roadway vehicles or equipment entering, exiting, or reversing inside the work area) by receiving audible, visual, haptic, or multi-modal alarms in their wearable devices. The application of wearable devices at work zones deserves further research efforts and field testing to understand the factors that motivate workers to adopt the technologies and that maximize their attention to the alarms. Choi et al. [*120*] found that perceived usefulness, social influence, and privacy are influential factors that must be considered to encourage construction workers to adopt wearable technologies. Based on initial testing at work zones recreated in virtual reality, Ergan et al. [*121*] found that workers' response to alarms from wearable sensors was sensitive to the alarm modality (vibration or visual) and frequency. The study called for more efforts that attempt to maximize workers' attention by changing the modality, frequency, and duration of the delivered alarms.

### *3.4.3   Vulnerable Road Users Guidance*

Reviewing the literature indicates that accommodating pedestrians and cyclists needs at work zones is an under-researched area. The Manual on Uniform Traffic Control Devices (MUTCD), in Chapter 6D, warns that pedestrians are generally reluctant to detour or to add additional distance to their trip [*122*]. Work zones create very complex settings for the pedestrians and cyclists to navigate. This complexity also exacerbates when pedestrian and cyclist detours keep changing according to the present construction staging on site. If pedestrians and cyclists are left with complex, unclear or unfeasible routing, they might be provoked and tempted to disregard signs and undertake risky travel paths. Traditional navigation or mapping systems do not provide high-resolution or updated routing to pedestrians or cyclists traversing complex work zone settings; they will have to stop and spend significant time understanding the posted signs and markings. However, the expected deployment of connected TCDs at WZs may not only facilitate CAV navigation, but they can also provide pedestrian and cyclist navigation systems with high-resolution and real-time detour data. Liao et al. [*123*] developed and tested a smart-phone navigation system that alerts visually impaired pedestrians when arriving at WZ and then guides them to traverse the construction area through audible messages. The system relied upon Bluetooth-connected beacons attached to light posts at the WZ site. The applicability of such systems can be easily extended to serve all pedestrians or cyclists and by using different guidance modalities (e.g., visual maps, audible messages).

As for CAVs interaction with pedestrians and cyclists, there is a well-established and growing literature covering this subject and it is beyond the scope of the present paper. However, it must be emphasized that all temporary pedestrian and cyclist detours should be provided with connected or machine-vision-detectable barricades or fencing systems so CAVs can be alarmed of these paths. This can be especially critical if these temporary paths encroach into the original vehicular travelled way, something that is commonly seen in tight urban construction sites.

### **3.5 WZ Data Systemization and Asset Monitoring**

With the rapidly evolving CAV technologies, WZs will be more connected and plenty of valuable data will become available and can be leveraged for several purposes. However, the data should be first standardized so it can be widely and efficiently used. Work zone data exchange (WZDx) is a recent effort launched by the US DOT and it aims to harmonize the collection, storage, and dissemination of work zone data by setting specifications and standards that should be followed locally and nationally [*124, 125*]. The WZDx complements the broader work zone data initiative (WZDI) which intends to manage work zone activity data (WZAD) by developing a data dictionary and implementation guidelines. This standardization should allow operators, owners, and third-party users to share work zone data easily and seamlessly. Examples of the data in WZDx include: WZ geometrics, locations of TCDs, workers' presence, start and end times of the WZ, speed limits, and others. Benefits of WZDx are not limited to vehicle navigation, transportation agencies and contractors alike can find several applications related to asset management and monitoring.

There are currently several methods to gather real-time data pertaining to a specific WZ. The simplest method is to require the construction crew to log in the system and insert or update the data manually [*115*]. The staff may further use a construction vehicle equipped with high accuracy GPS device [*126*], the vehicle travels in the first open lane near the work zone and the staff records the location of key WZ features while travelling (beginning and end of WZ, beginning and end of workers' area, etc), the collected data are then uploaded to the back office. Using connected TCDs can be a more efficient and dynamic method. Knickerbocker et al. [*89*] used smart arrow boards that transmit the location of the start and the end of the work zone on several sites in Iowa. Parikh et al. [*90*] used "asset tracking devices" that can be attached to different types of TCDs (e.g., cones, signs, barrels) and then track and update their real-time location and movement. It is worth noting that all these studies [*126, 89, 90*] used GPS-powered localization and the accuracy of the TCD location remains a challenge (e.g., $\pm$ 8 m [*126*] and $\pm$ 50 ft [*89*]).

Such data resolution may only support driver's awareness of the WZ presence ahead. If lane-level information is made available, CAVs may tolerate longitudinal geometric inaccuracy by deciding to join the open lanes early, they however still need high-resolution in the cross-sectional data to avoid striking the barrier or the TCDs delimiting the open path. More advanced localization techniques that provide higher resolution of the WZ geometry are evidently needed to support high automation levels while also serving the WZDx other purposes.

With connected WZ data, transportation agencies will be able to respond to many challenges related to asset monitoring. Due to several reasons, the actual on-field WZ location may differ than the planned or announced location, e.g., Knickerbocker et al. [*89*] found that the average difference was 0.96 mi over 18 WZs in Iowa. Weather conditions and equipment unavailability may also force contractors to cancel or reschedule work activities. The real-time and remote exchange of WZ data will resolve the frequent discrepancies in WZ time and location. In addition, with connected TCDs, verification of WZ geometrics can be done remotely reducing the time and human resources needed by the inspection agency. Inspection algorithms based on national manuals such as the MUTCD can be developed and used to verify the spacing and the types of the used TCDs [*90*]. For extra scrutiny or in the absence of connected TCDs on site, inspection agencies may also use their own connected "inspection" vehicles. Mekker et al. [*127*] used a LiDAR-equipped vehicle to verify that work zone geometry adheres to the standards, the vehicle was able to inspect long directional miles quickly and efficiently without the need for the inspection staff to leave the vehicle. The remote verification of WZ data should allow for more reliable monitoring of WZs that have been traditionally overlooked or not continuously checked, e.g., mobile, short-term, rolling, and alternating WZs which shift quickly from one side of the roadway to the other. TCDs that are placed behind emergency vehicles and incident management crew can also be similarly connected and inspected remotely. Exchanging WZ data should also assist agencies in monitoring the project contract and schedule especially when the WZ occupies sensitive areas or uses lane-rental systems. The WZ data can be further used post-construction for auto accident forensics to decide who was at fault and apportion liability fairly.

On the other hand, the contractor can use the connected WZ data to track the movement of their equipment and devices when shifted from one site to another, maximize their utilization, and better manage their use and maintenance schedules. The connected TCDs can remotely alert the site operator or supervisor if they were struck by a vehicle or moved by other reasons (e.g., weather conditions or wind). If the connected TCDs become also automated (*Section 3.2.2*), the contractor can increase productivity, reduce person-hours needed, and add more shifts especially in the cases of rolling and alternating closures that are typical for lane marking and pavement resurfacing works. Archived WZ data records can be used for performance evaluation and research by the contractor organization.

## 4. REGULATORY CONSIDERATIONS

### 4.1 WZs in Some Existing CAV Regulations

Existing regulations and guidelines are not silent about the interaction between CAVs and WZs. In an early initiative to regulate the use of CAVs, the US federal government [*128*] defined an operation design domain (ODD) based on the available object and event detection and response (OEDR) competencies. The regulation states clearly that within an ODD, the CAV's OEDR should be able to deal with temporary work zones. Listed examples of OEDR competencies relevant to WZs included: detection of work zones, responding to temporary traffic control devices, and following police or construction workers when manually directing traffic. Without these competencies, the ODD of CAVs will be limited to highway segments not having WZs. A new edition of the MUTCD is underway, and the FHWA published a notice of proposed amendments (NPA) for this edition which will include a fifth part dedicated to addressing automated vehicles' needs [*129*]. To specifically support the vehicle's machine-vision at WZ areas, the NPA asked to enhance visibility of channelizing devices, use uniform line width, obliterate markings that are no longer valid, minimize pavement scarring, and stop painting over existing markings with black paint or spray.

Review of the existing guidelines concerning CAV operations at WZs reveals that the guidelines are still very general and at the functional level. The NPA of the forthcoming MUTCD only considered broadly the machine vision of automated vehicles at WZs, other technologies like communication, connectivity, and smart devices were not discussed. The generality of the guidelines is advantageous at this stage because it keeps the regulations flexible to adapt to new technologies especially those that are not yet matured or fully tested, and it avoids putting restrictions on CAV manufacturers who are still at the innovation phase.

Future guidelines should also consider amending traffic management plans (TMPs) required by contractors to obtain the WZ permit. The TMPs should provide CAV impact assessment and state how the WZ will serve or detour CAVs; also, the contractor should be in compliance with WZ data sharing programs such as the WZDx (*Section 3.5*).

### 4.2 Cross-border Harmonization

Driver's and vehicle's licenses can typically be used in different states or countries based on reciprocity agreements. According to Fagnant et al. [*130*] and Smith et al. [*131*], existing laws need to better clarify how and when these reciprocity agreements are extended to cover CAV licensing. Because WZ operations require the use of a unique set of TCDs, states and countries may need to standardize these devices and their technologies to preserve the reciprocity of CAV licenses; otherwise, CAV navigation may become unreliable across states or countries.

Existing guidelines already call for uniform TCDs across states or provinces. For example, the FHWA requires all states to be in substantial conformance with the national MUTCD, only limited supplements can be developed by the states and these also must be reviewed and approved by the FHWA [*132*]. Most European countries also adopt the Vienna Convention on Road Signs and Signals [*133*]. To maintain such uniformity in national manuals, states or provinces must coordinate to adopt similar WZ devices and technologies, the top or federal government should also regulate the importing or manufacturing of CAVs to ensure they all have the needed types of sensors and technologies and their specifications. However, international harmonization of WZ technologies and TCDs seems to be more complicated and challenging. A quick solution to cross-border interoperability in the short-term is to require "out-of-state" or "out-of-the-country" CAVs to switch off automation when approaching a WZ. However, this requires enforcement and reliable communication and may also become impractical or inconvenient in some urban areas where construction activities are common fixtures constituting a significant proportion of the network. International harmonization is more needed across borders that are traditionally flexibly crossed by drivers, e.g., between Canada and USA, between EU countries, etc.

### 4.3 Insurance and Liability

Insurance and liability are among other major challenges for CAVs at WZs. Liability of CAV accidents is a complex problem [66, 134], most jurisdictions have not yet developed liability rules for CAV crashes and only few have prepared supplementary provisions [134]. With fault liability, the vehicle manufacturer will be liable if the crash occurs under automation or due to manufacturing deficiencies whereas the driver will be liable if the crash was due to driver's error or refusal to retake vehicle control. However, it is evident that the cause of many crashes can be difficult to identify. Pattinson et al. [66] also highlighted that conveying certain messages or consents to the driver throughout the vehicle digital interface while driving may not constitute a solid legal ground to hold the driver liable.

Liability at WZs can be a more complicated question due to several reasons. First, apportioning liability is not limited only to the driver, manufacturer, or owner of the vehicle, the contractor's presence on site adds another important party. Fault liability regime, for example, should hold the contractor liable if they failed to provide standardized TCDs or communication technologies that accommodate CAV systems. Second, if a CAV accident becomes unavoidable near a WZ, many legal questions arise such as how CAVs should behave to minimize the damage? Should CAVs avert from the construction area, or should they alternatively avert from the roadway? What types of data the contractor must share with the approaching vehicles in a real-time manner to make these decisions more informed (e.g., equipment and number of workers on site, their exact locations)?

The insurance premiums of CAVs may rise in the short run because repair and replacement of advanced technologies can be costly at early market penetration and because insurers need more empirical evidence that supports CAV competencies. The premiums should however decline in the long run when the cost of CAV systems becomes affordable and when the technology matures and proves its ability to significantly lower crash rates [135]. Nevertheless, insurers of CAVs are expected to dispute lowering CAV premiums and be concerned about complex highway settings that challenge CAV systems, WZs are evidently legitimate examples of these worrying situations. A crash within a connected WZs can be very costly, it may require repair and replacement of expensive devices including not only CAV systems but also the contractor instruments (e.g., smart cones and barrels as compared to traditional TCDs). In addition, because apportioning liability at WZs is a difficult case, litigation will be a lengthy and complex process and the insurers will endure extra time and cost to process the claims of WZ crashes. These concerns are not limited to CAVs' insurer, they can also worry the contractor's insurer. Insurers therefore must be convinced about how safe and reliable CAV systems are when traversing WZs and the litigation process must be efficient so claims can be handled promptly. Sharing pre-crash and post-crash WZ data through governmental systems (e.g., through WZDx programs) should support crash forensics and make the claim process quicker. Developers of CAV liability regimes and insurance policies should consider adding special provisions and rules that address CAV crashes at WZs. Insurers are among the key stakeholders that must be engaged to successfully harmonize CAVs at WZs.

### 4.4 Privacy

CAV data ownership, accessibility, and uses are critical issues that require further definition by legislatures [130]. Connected WZs expand privacy concerns as the contractor becomes an additional party that can own or access the data. When a communication is established between CAVs and the contractor devices, the vehicle and the contractor can obtain critical data about each other. Privacy concerns at WZs are tightly related to liability. The drivers, CAV manufacturers, and the contractors should be concerned about "what data can be accessed by each party" and how this can be used against them in a post-crash litigation process. Several privacy-related questions must therefore be investigated such as what type of data the vehicle and the contractor can store, use, or share at WZs? Should some types of data be stored at a third-party or a governmental agency (e.g., through WZDx programs)? Workers and pedestrians may object to detailed surveillance and face-recognition systems which may consequently challenge many automation-based safety functions. Privacy concerns at WZs must be deliberately considered and the government should establish regulations that address the trade-off between the benefits of sharing data and the privacy constraints.

## 5. RESEARCH AGENDA

The preceding sections highlight that the existing literature treating CAVs at WZs is still at its early emergence, there are many concepts, technologies, and considerations yet to be explored or better understood. This section provides a research agenda for deploying CAV systems at WZ settings. Two main subjects are discussed: outlining a research framework structure and identifying the major research needs. Experts in the field of CAVs and WZs were surveyed in order to verify and enrich the proposed research directions.

### 5.1 Research Framework

Figure 9 outlines a proposed framework for researching CAV systems at WZs. Figure 9(a) shows the technical components and the influential factors to be explored. The ripple effect diagram was used to demonstrate the sequential implications of the vehicular-related subjects. Technologies are placed in the centre of the ripple and they mainly include automation, connectivity, and smart devices. These technologies impact each other, for example, vehicle automation may require the use of a smart TCD which also requires the use of communication of specific latency and range. Other types of technologies may also be added to the ripple centre. The first-order implications are represented by the behaviours of CAV systems, drivers of CAVs, and drivers of conventional vehicles, these behaviours are directly influenced by the available technologies and TCDs. In a mixed traffic WZ, the three behaviours will affect each other. Understanding the behaviours of the vehicles and the drivers (e.g., conservatism, assertiveness, aggressiveness, etc) will make it possible to parametrize car-following and lane-changing algorithms which represent the second-order implications. These algorithms along with the MPR will finally generate the third-order implications which are the ultimate mobility, safety, and environmental measures. Many external non-vehicular factors may also impact the ripples. The order of the ripples can assist in prioritizing the research needs. For example, investing more in developing technologies (ripple centre) and understanding drivers' behaviour (first-order implications) should reduce the need for making many assumptions when conducting mobility or safety studies (third-order implications) and should offer more solidified research bases.

Figure 9(b) summarizes the regulatory context and who are the engaged key stakeholders and how they should interact in the research and development process. The leading stakeholder is represented by the governmental transportation agencies who should establish and set the major research guidelines and standards to be followed by all stakeholders. However, transportation agencies should integrate the perspectives and the views of various stakeholders, who work directly in the field and better understand the actual implementation needs, by implementing stakeholder engagement plans. By keeping such engagement plans on-going and updated, the gap between research and actual implementation should be narrowed and better controlled.

Because many CAV technologies have not yet matured, the government should update the guidelines periodically based on the research results and technology advancement. In the early stages, the guidelines should be relaxed to keep the doors open for innovation and testing new ideas. With time, some technologies will fade and be restricted, and others will mature and be adopted. Therefore, more sustainable and restrictive standards should be adopted in the advanced stages to narrow down the research directions. All levels of government (top, middle, and local) are in the research loop. Figure 9(b) associates each level of government with a set of different pertinent responsibilities; these were drawn from the recent literature in North America and Europe [e.g., 128, 134, 136]. When it comes to WZs, the most important task is to ensure national cross-border harmonization of technologies and standards between states or jurisdictions, and this should be undertaken by the top-level of government. Each registered equipped-vehicle should ensure traversing each equipped-WZ compatibly.

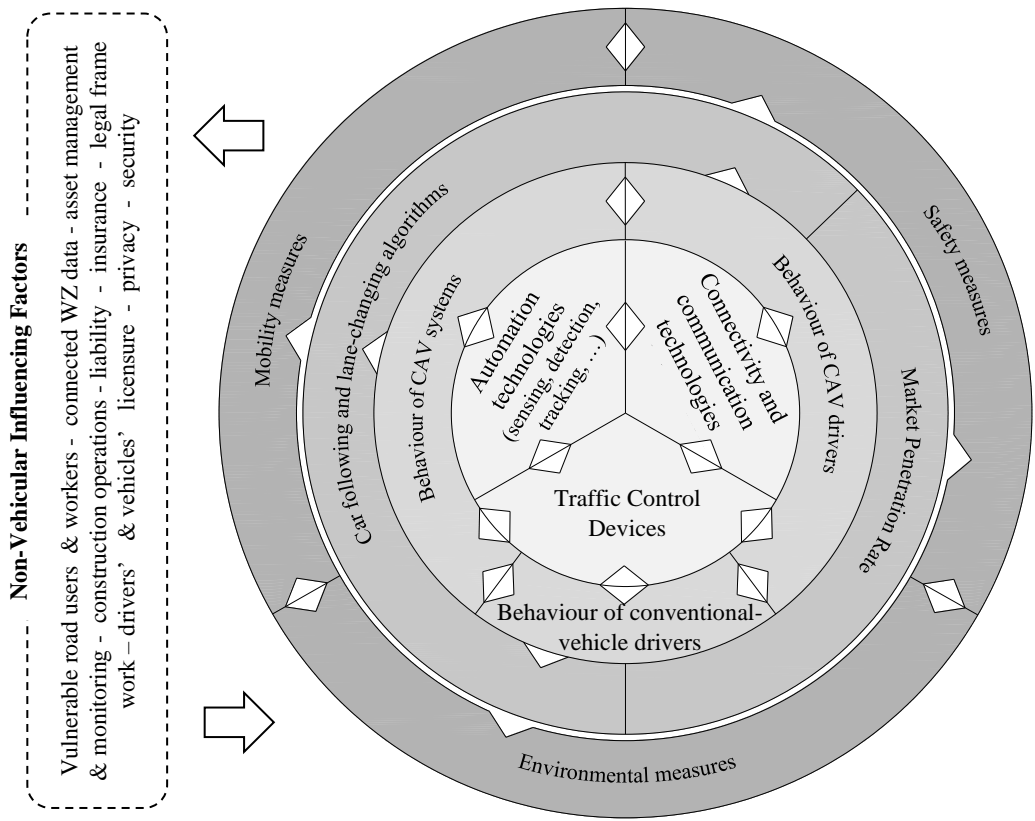

**(a)** The technical components and their relationship

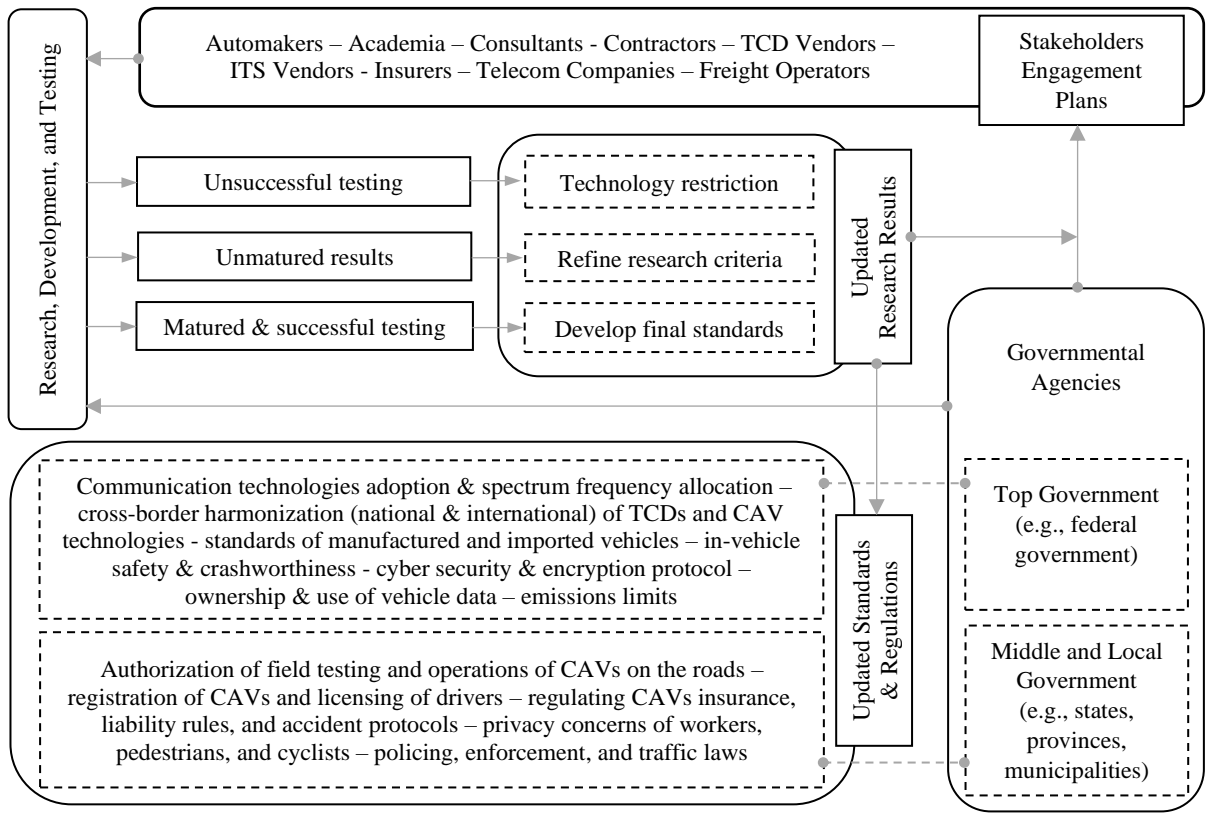

**(b)** The regulatory process and structure

**Figure 9. Framework for researching CAV systems at WZs.**

**5.2 Research Needs**

Identification of the most pressing research needs pertaining to the coexistence of CAVs and WZs was done through two steps. First, a list of proposed research topics was prepared based on the status of available literature and the major knowledge gaps found in the present research. The list was also rectified through the peer-review process. Second, a survey was distributed to experts in the fields of CAVs and WZs asking them to: (i) rate the list of research topics identified in the first step, and (ii) add additional topics if the list was missing important ones. The experts were selected carefully so that they: (i) represent both academia and research-oriented industrial organizations, and (ii) have a robust knowledge in CAV systems as demonstrated by their scholarly publications or by the objectives, activities, and achievements of their organizations. Examples of the surveyed industrial organizations include relevant transportation agencies and Intelligent Transportation System (ITS) associations and societies in Canada, USA, and Europe. 31 experts from North America, Europe, and Asia filled the survey, 16 from academia and 15 from the industry.

Table 4 outlines the finalized list of research needs. The topics, nineteen in total, were classified into five broad research themes: mobility and Safety, environment, driver's behaviour, technologies and infrastructure, and regulations. For each topic, Table 4 further provides a description, the status of existing literature, and the experts survey rating. Those topics without ratings are the ones added after the survey considering the recommendations given by the survey respondents. Only three additional topics were added to the initial list as many of the topics recommended by experts were actually a reiteration of the original list of topics. Appendix A provides in detail the survey rating results. The rating scale is a Likert-type with the following five points (levels): 1 (very low importance), 2 (low importance), 3 (average importance), 4 (high importance), and 5 (very high importance). All ratings exceeded 3.0 indicating above average-importance for all topics. Topics with relatively higher ratings, i.e., near or above 4.0, included drivers' behaviour (topics 9 and 10), vulnerable road users and workers (topic 8), crash analysis (topic 2), and several technological aspects (topics 13, 14, and 15). The ratings generally accord with the ripple effect diagram in Figure 9(a) and its implicated priority because a relatively higher rating was given to the topics at the ripple centre (i.e., technologies) and the first-order implications (i.e., driver's behaviour).

Experiments that can be used for searching the topics in Table 4 include simulation and field-testing on testbeds or on real-world highways, i.e., when CAVs become operating on the road network. Several new search methods have been used recently that are a combination of several tools including: (i) integrated driving simulation [*e.g., 137-139*] whereby the driving simulator connects to a microscopic traffic simulation software which generates the virtual traffic and highway environment, (ii) Vehicle-In-The-Loop [*e.g., 140, 141*] whereby the self-driving vehicle moving on the testbed surface connects to a microscopic traffic simulator, and (iii) naturalistic driving experiments [*e.g., 142, 143*] whereby the automated or connected vehicle moves on real-world highways and is equipped with cameras and sensors that measure vehicle performance, driver behaviour and the surrounding conditions. As mentioned earlier, the literature needs more field-testing and real-world studies. Nevertheless, simulation tools should still play a major role because they can flexibly analyze a multitude of traffic and highway variables and they can also narrow down the scope of the field experiments and their associated cost.

**Table 4. Research Needs for Deploying CAV Systems at WZs**

| No | Topic | Theme | Description | Status in Existing Literature | Experts Survey Rating |
|---|---|---|---|---|---|
| 1 | **Impact of CAVs on WZ capacity and travel time** | Mobility and Safety | How to build capacity models that incorporate clustered vehicles, cooperative lane change, dedicated CAV lane, and different MPRs? | Under-researched | 3.65 |
| 2 | **Impact of CAVs on crash types and frequency at WZs** | Mobility and Safety | What are the crash modification factors that can be achieved due to CAV-powered concepts (e.g., automation, crash warnings, speed harmonization, etc)? | Under-researched | 3.97 |
| 3 | **Clustering in the WZ vicinity** | Mobility and Safety | When and where clustering should be encouraged, permitted, or prohibited at WZs? How clustering strategies can be influenced by each travel lane, WZ configuration, MPR, and the distance remaining to the WZ taper? | Not researched | 3.69 |
| 4 | **Cooperative lane change at WZs** | Mobility and Safety | How to deploy the cooperative lane change concepts at WZs while also considering urgency of the lane change request, WZ configuration, traffic volume, MPR, and communication latency? | Under-researched | 3.67 |
| 5 | **Smart lane flow distribution at WZs** | Mobility and Safety | How to define a grouping-based lane change technique whereby drivers change their lanes progressively, i.e., group-after-group, to avoid an all-at-once lane change disruption and in order to achieve an ideal lane flow distribution upstream of the WZ? | Not researched | 3.52 |
| 6 | **Connected and automated heavy vehicles at WZs** | Mobility and Safety | How will heavy vehicles be treated in a mixed traffic WZ? What specific considerations must be given to connected and automated trucks? Where and when truck platooning should be permitted? | Not researched | 3.77 |
| 7 | **How to improve existing smart WZ systems with low CAV MPR?** | Mobility and Safety | Examining how existing off-the-shelf smart WZ systems (e.g., queue warning, variable speed limit, alternate route advisory, etc) can benefit from CAV technologies and their data in a mixed traffic. | Under-researched | 3.78 |
| 8 | **Vulnerable road users and workers at connected WZs** | Mobility and Safety | What are the specific challenges and opportunities for pedestrians and cyclists when they traverse connected WZs and workers when they present on-site? What technology-based measures can be used and developed to enhance their mobility and safety? | Under-researched | 4.10 |
| 9 | **Automated enforcement at WZs** | Mobility and Safety | How can enforcement techniques benefit from connectivity and automation at WZs? How driver's behaviour, safety, and mobility change if the real-time speed is monitored remotely and if the speed limit can be enforced inside the vehicle through connectivity? | Not researched | Unrated* |
| 10 | **Driver behaviour of human-driven vehicles in a mixed traffic WZ** | Driver's Behaviour | Would human-driven vehicles maintain an anxiously conservative gap with nearby CAVs or drive more assertively? Would they imitate the behaviour of CAVs without being connected? Would they cooperate? | Not researched | 4.21 |

(*) these topics were not rated because they were added after the survey based on the experts' feedback and advice.

1
2
**Table 4 (continued). Research Needs for Deploying CAV Systems at WZs**

| No | Topic | Theme | Description | Status in Existing Literature | Experts Survey Rating |
|---|---|---|---|---|---|
| 11 | **Driver behaviour of CAVs in a mixed traffic WZ** | Driver's Behaviour | Can drivers of CAVs accept a tightened following gap in a complex and unpredictable WZ geometry or would they alternatively behave conservatively and maintain long gaps? | Not researched | 3.94 |
| 12 | **CAV environmental impact at WZs** | Environment | How much CAV-powered environmental functions such as eco-driving, platooning, cooperative lane change, eco-routing, and enhanced capacity can reduce emissions at WZs? | Not researched | 3.55 |
| 13 | **Applicability of different vehicle automation levels at WZs.** | Technologies and Infrastructure | What are vehicle automation levels that can be attainable at WZ areas? How automation level may vary by sub-area (i.e., early upstream, approach area, queuing area, activity area, and termination area)? | Not researched | 3.90 |
| 14 | **Suitability of different communication techniques at WZs.** | Technologies and Infrastructure | What are the best communication technologies by application (DSRC, 5G, satellite, etc)? How some factors may influence the selection (e.g., short vs. long term, rural vs. urban, congested vs non-congested WZs, cost, coverage, latency, security, MPR, etc)? Would a hybrid approach be favourable? | Moderately researched and growing | 3.97 |
| 15 | **Deployment of connected TCDs and specific types of technologies at WZs.** | Technologies and Infrastructure | What are the needed functions and specifications of smart and connected TCDs at WZs? What are the possible applications of specific types of technologies at WZs (e.g., drones, real-time tracking, cloud and edge computing, detection, etc)? | Moderately researched and growing | 3.82 |
| 16 | **Construction operations at connected WZs** | Technologies and Infrastructure | How robotics and smart devices can reduce person-hours and maximize site productivity? How connected WZs can support construction operations and the tracking of devices, workers, and equipment? | Under-researched | 3.41 |
| 17 | **Standardization and systemization of WZ data sharing** | Technologies and Infrastructure | Developing specifications, standards, and framework to facilitate data sharing between infrastructure owners and operators, CAV developers, and third-party travel information system providers. What are the needs to better harmonize and standardize sharing WZ activity data nationally? What are the applications of WZ data other than CAV navigation (e.g., post-crash forensics, asset monitoring, etc)? | Moderately researched and growing | Unrated* |
| 18 | **Legal and insurance issues at connected WZs** | Regulations | Are the insurance industry and the legal system ready to treat CAV operations at WZs? What changes are needed? What extra complexity may result when assigning liabilities to vehicles, drivers, pedestrians, workers, and the contractor in a connected WZ context? | Not researched | 3.77 |
| 19 | **Education, licensing, and work permit needs** | Regulations | What education programs are needed to ensure efficient CAV operations? Should CAVs' registration and drivers' licensing consider some specific technologies and driving skills at connected WZs? What extra requirements should be added to traffic management plans prepared by contractors to obtain work permits at connected WZs? | Not researched | Unrated* |

3 (*) these topics were not rated because they were added after the survey based on the experts' feedback and advice

# 6. CONCLUSIONS

The near future will bring the coexistence of CAVs and WZs. Integrating CAV technologies and concepts into WZ settings has therefore become an important research direction that is still emerging and raising many interesting knowledge gaps. This paper provides a state-of-the-art review on this subject aiming to assist in paving the way forward. The paper presented the main concepts, challenges, and opportunities of deploying CAVs at WZs, reviewed the literature, and outlined a research agenda. The following are the main conclusions of the paper.

CAVs have the opportunity to solve many traffic challenges at WZs as compared to conventional vehicles. Around 18 concepts and functions of CAV systems related to mobility, safety, and environment were reviewed and discussed. The paper proposed a high-level spatial distribution of these functions and concepts along the WZ area which was subdivided into five segments: further upstream, approach area, queuing area, WZ activity, and termination area. Nevertheless, achieving CAV benefits at WZs faces challenges. First, driver's behaviour towards connectivity and automation at WZs is still not well researched and can significantly alter the efficiency of many CAV functions. Second, mobility, safety, and environment measures may conflict; enhancing all measures at once is not always possible. Third, many of the reported CAV benefits are highly dependable on technology advancement and maturity and the presence of very high MPR. Finally, excessive use of CAV functions may return adverse impacts or unintended consequences in some circumstances, e.g., excessive CAV rerouting away from WZs may increase the average trip distance and thereby instigate exposure to other hazards.

Existing research findings indicate that vehicles can detect signs, pavement marking, and other traffic control devices using computerized vision-based techniques (camera-based and LiDAR); however, detection success rate remains a challenge calling for more confirmatory research efforts. On the other hand, communication-based detection provides a more reliable approach but the TCDs need to be supplied with tags or sensors. Nevertheless, some connected TCDs are in use today and their market is expected to grow.

The connectivity of WZ opens the doors for many traffic and non-traffic applications. CAV drivers, governmental agencies, and owners and operators of construction sites can all benefit from WZ data connectivity especially if harmonized and systemized through a national data exchange program. Workers can also benefit from wearable technologies to enhance their safety and vulnerable road-users can use real-time high-resolution guidance systems to easily navigate through complex WZs.

Regulatory challenges for CAV deployment at WZs are not less important than the technical ones. Plenty of work has yet to be done to set regulations that address cross-border harmonization, license reciprocity of vehicles and drivers, amendments to national traffic manuals, requirements of construction permits and traffic management plans, liability, insurance, and privacy concerns.

The paper proposed a research framework and identified specific research needs for traffic agencies and prospective researchers who are interested in searching the coexistence of CAVs and WZs. The research needs were rated and rectified based on a survey of 31 experts in the area of CAVs and WZs. Research topics that received relatively higher ratings include driver's behaviour, vulnerable road users and workers, crash analysis, and advancing specific technologies of automation, communication, and TCDs. A lot of research work has yet to be conducted in order to better harmonize the forthcoming coexistence between CAVs and WZs.

**CRediT Authorship Contribution Statement**

**Amjad Dehman:** Conceptualization, Methodology, Investigation, Writing – Original Draft. **Bilal Farooq:** Methodology, Supervision, Writing – Review and Editing, Project Administration, and Funding acquisition.

**Acknowledgement**

This research was funded and supported by Government of Canada and Lazaret Capital Inc. through the Mitacs Elevate fellowship program.

# Appendix A. Expert Survey Results: Rating of Proposed Research Topics

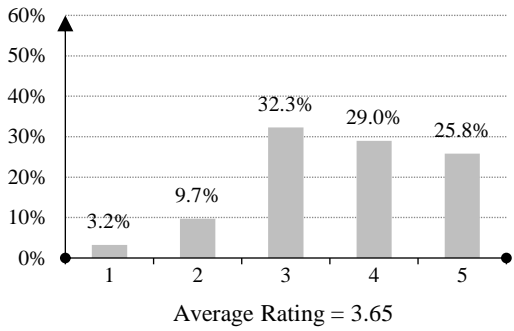

**Topic No 1: Impact of CAVs on WZ capacity and travel time**
Average Rating = 3.65

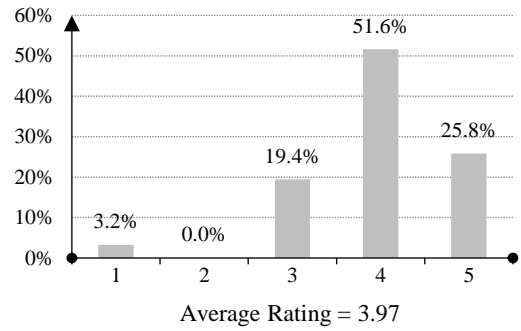

**Topic No 2: Impact of CAVs on crash types and frequencies at WZs**
Average Rating = 3.97

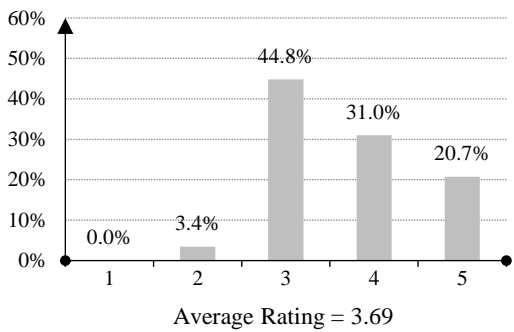

**Topic No 3: Clustering in the WZ vicinity**
Average Rating = 3.69

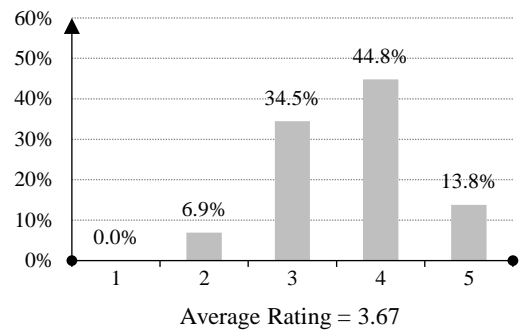

**Topic No 4: Cooperative lane changes at WZs**
Average Rating = 3.67

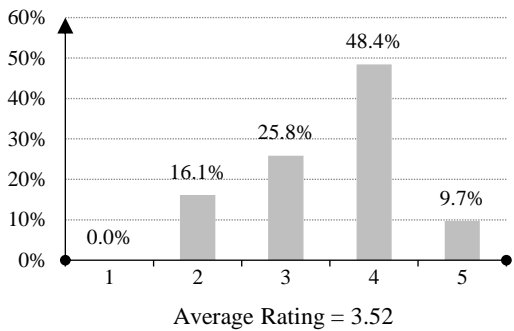

**Topic No 5: Smart lane flow distribution at WZs**
Average Rating = 3.52

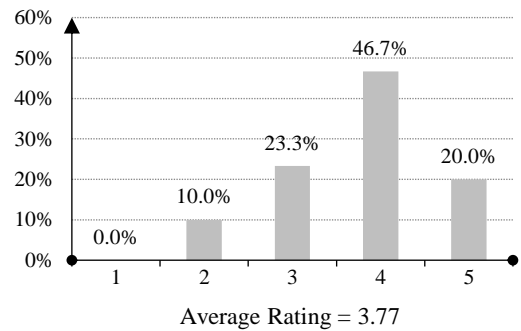

**Topic No 6: Connected and automated heavy vehicles at WZs**
Average Rating = 3.77

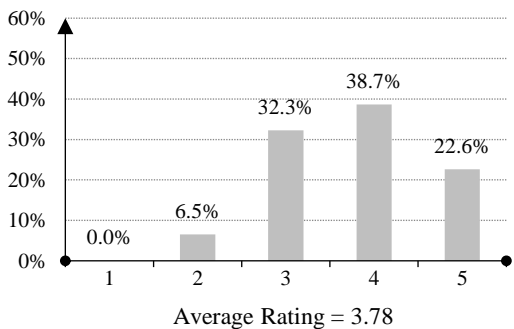

**Topic No 7: How to improve existing smart WZ systems with low MPR?**
Average Rating = 3.78

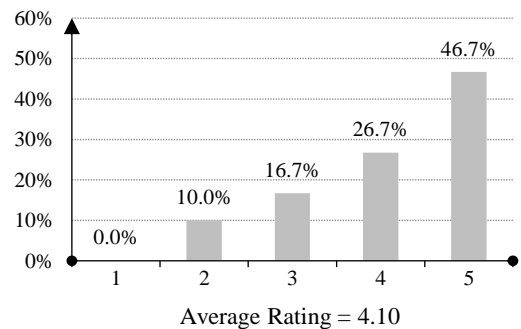

**Topic No 8: Vulnerable road users and workers at connected WZs**
Average Rating = 4.10

**Note:** The rating scale is a Likert-type with the following five points (levels): 1 (very low importance), 2 (low importance), 3 (average importance), 4 (high importance), and 5 (very high importance). Topics 9, 17, and 19 were unrated and added post the survey.

**Appendix A (continued). Expert Survey Results: Rating of Proposed Research Topics**

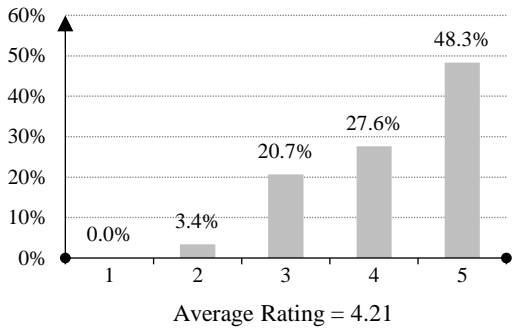

Average Rating = 4.21

**Topic No 10: Driver behaviour of human-driven vehicles in a mixed traffic WZ**

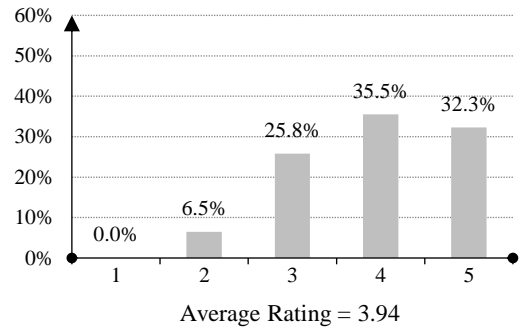

Average Rating = 3.94

**Topic No 11: Driver behaviour of CAVs in a mixed traffic WZ**

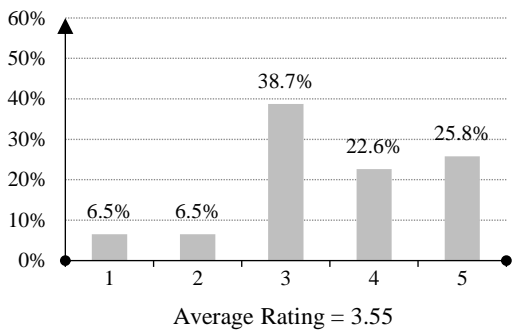

Average Rating = 3.55

**Topic No 12: CAV environmental impact at WZs**

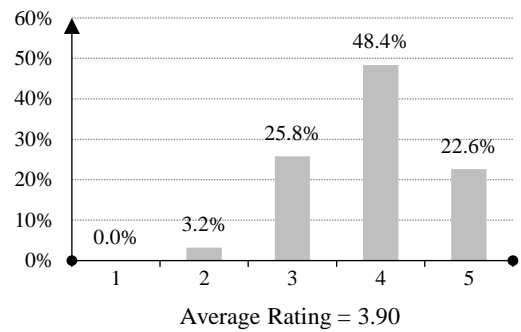

Average Rating = 3.90

**Topic No 13: Applicability of different vehicle automation levels at WZs**

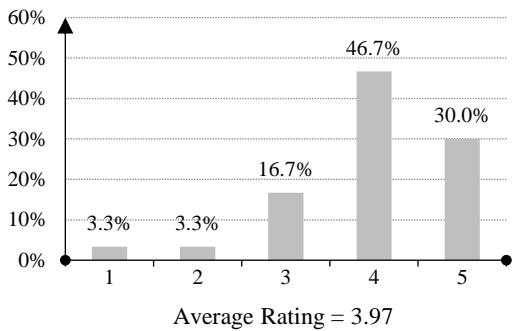

Average Rating = 3.97

**Topic No 14: Suitability of different communication techniques at WZs**

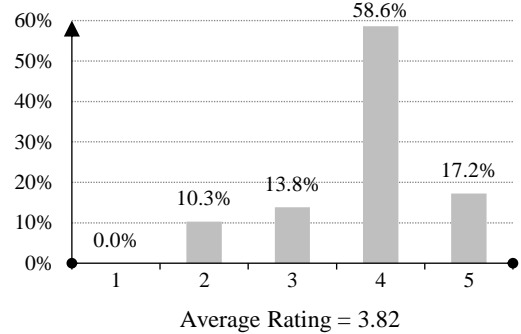

Average Rating = 3.82

**Topic No 15: Deployment of connected TCDs and specific types of technologies at WZs**

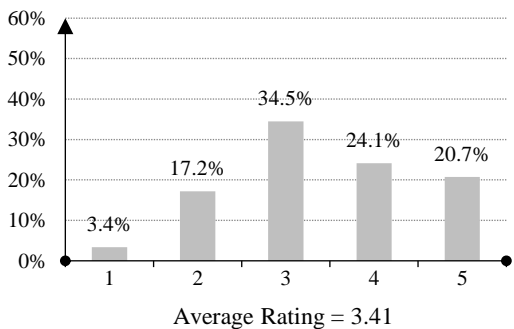

Average Rating = 3.41

**Topic No 16: Construction operations at connected WZs**

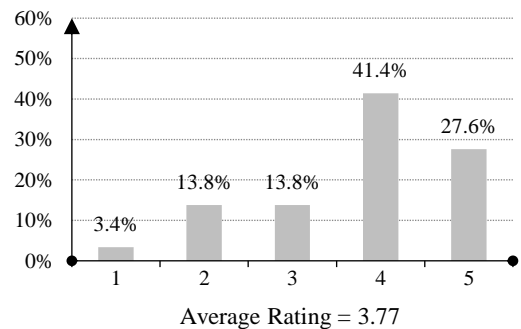

Average Rating = 3.77

**Topic No 18: Legal and insurance issues at WZs**

**Note:** The rating scale is a Likert-type with the following five points (levels): 1 (very low importance), 2 (low importance), 3 (average importance), 4 (high importance), and 5 (very high importance). Topics 9, 17, and 19 were unrated and added post the survey.

Dehman and Farooq  41103. Xu, Z., X. Li, X. Zhao, M. H. Zhang, and Z. Wang. DSRC versus 4G-LTE for Connected Vehicle Applications: A Study on Field Experiments of Vehicular Communication Performance. Journal of Advanced Transportation, Volume 2017, Article ID 2750452, 2017, pp. 1-10.
104. Barros, M. T., G. Velez, H. Arregui, E. Loyo, K. Sharma, A. Mujika, and B. Jennings. CogITS: Cognition-Enabled Network Management for 5G V2X Communication. IET Intelligent Transport Systems, Volume 14, Issue 3, 2020, pp. 182-189.
105. Ansari, S., J. Ahmad, S. A. Shah, A. K. Bashir, T. Boutaleb, and S. Sinanovic. Chaos-based Privacy Preserving Vehicle Safety Protocol for 5G Connected Autonmous Vehicle Networks. Transactions on Emerging Telecommunications Technologies. Volume 31, Issue 5, 2020.
106. Do, D-T., T-TT. Nguyen, C-B. Le, and J. W. Lee. Two-Way Transmission for Low-Latency and High-Reliability 5G Cellular V2X Communications. Sensors, Vol. 20, Issue 2, 2020.
107. Wang, P., B. Di, H. Zhang, K. Bian and L. Song. Platoon Cooperation in Cellular V2X Networks for 5G and Beyond. IEEE Transactions on Wireless Communications, Vol. 18, No. 8, 2019, pp. 3919-3932.
108. Ge, X., H. Cheng, G. Mao, Y. Yang, and S. Tu. Vehicular Communications for 5G Cooperative Small-Cell Networks. IEEE Transactions on Vehicular Technology, Vol. 65, No. 10, 2016, pp. 7882-7894.
109. Reid, T. G. R., A. M. Neish, T. Walter, and P. K. Enge. Broadband LEO Constellation for Navigation. Journal of the Institute of Navigation, Vol. 65, No. 2, 2018, pp. 205-220.
110. Ge, H., B. Li, L. Nie, M. Ge, and H. Schuh. LEO constellation optimization for LEO enhanced global navigation satellite system (LeGNSS). Advances in Space Research, Volume 66, Issue 3, 2020, pp. 520-532.
111. Iannucci, P. A., and T. E. Humphreys. Fused Low-Earth-Orbit GNSS. arXiv:2009.12334. 2020, version v1.
112. Reid T. G. R., B. Chan, A. Goel, K. Gunning, B. Manning, J. Martin, A. Neish, A. Perkins, and P. Tarantino. Satellite Navigation for the Age of Autonomy. Proceedings of the 2020 IEEE/ION Position, Location and Navigation Symposium (PLANS), Portland, OR, USA, 2020, pp. 342-352.
113. Zaman, A. U., M. I. Hayee, N. Katta, and S. Mooney. Traffic Information System to Deliver In-Vehicle Messages on Predefined Routes: Use of Dedicated, Short-Range Vehicle-to-Vehicle Communication. In Transportation Research Record: Journal of the Transportation Research Board, No. 2559, 2016, pp. 73-80.
114. Ibrahim, U., M. I. Hayee, and E. Kwon. Hybrid Work Zone Information System with Portable Changeable Message Signs and Dedicated Short-Range Communication. In Transportation Research Record: Journal of the Transportation Research Board, No. 2380, 2013, pp. 29-35.
115. Azadi, F., Y. Adu-Gyamfi, C. Sun, and P. Edara. Mobile Applications Development and Testing for Work Zone Activity Real-Time Data Collection. In Transportation Research Record, Journal of the Transportation Research Board, Vol. 2674(6), 2020, pp. 351-362.
116. Mollenhauer, M., E. White, and N. Roofigari-Esfahan. Design and Evaluation of a Connected Work Zone Hazard Detection and Communication System for Connected and Automated Vehicles (CAVs). Prepared by Safe-D National UTC and Virginia Tech Transportation Institute for the US Department of Transportation. 2019. <https://safed.vtti.vt.edu/wp-content/uploads/2020/07/03-050_FinalResearchReport_Final.pdf>
117. Malveaux, C., M. De Queiroz, X. Li, H. Hassan, and Z. He. Real-Time Work Zone Traffic Management via Unmanned Air Vehicle. Prepared by Transportation Consortium of South-Central States (Tran-SET) for the US Department of Transportation. 2020. <https://digitalcommons.lsu.edu/cgi/viewcontent.cgi?article=1068&context=transet_pubs>
118. Kim, K., S. Kim, and D. Shchur. A UAS-based Work Zone Safety Monitoring System by Integrating Internal Traffic Control Plan (ITCP) and Automated Object Detection in Game Engine Environment. Automation in Construction, Vol. 128, 2021.
119. Sun, H., Z. Zhang, R. Q. Hu and Y. Qian. Wearable Communications in 5G: Challenges and Enabling Technologies. In IEEE Vehicular Technology Magazine, Vol. 13, No. 3, 2018, pp. 100-109.
120. Choi, B., S. Hwang, and S. Lee. What Drives Construction Workers Acceptance of Wearable Technologies in the Workplace? Indoor Localization and Wearable Health Devices for Occupational Safety and Health. Automation in Construction, No. 84, 2017, pp. 31-41.
121. Ergan, S., Ozbay, K., S. D. Bernardes, Z. Zou, and Y. Shen. Increasing Work Zone Safety: Worker Behavioral Analysis with Integration of Wearable Sensors and Virtual Reality. U.S. Department of Transportation, University Transportation Centers Program. 2020. <https://c2smart.engineering.nyu.edu/wp-content/uploads/69A3551747124-Increasing-Work-Zone-Safety-Worker-Behavioral-Analysis-with-Integration-of-Wearable-Sensors-and-Virtual-Reality.pdf>
122. Manual on Uniform Traffic Control Devices for Streets and Highways. 2009 Edition. US Department of Transportation, Federal Highway Administration.
123. Liao, C. F. Development of a Navigation System Using Smartphone and Bluetooth Technologies to Help the Visually Impaired Navigate Work Zones Safely. Prepared by Department of Civil Engineering at University of Minnesota for Minnesota Department of Transportation, Report No. MN/RC 2014-12, 2014. <https://www.lrrb.org/pdf/201412.pdf>
124. U.S. DOT. Work Zone Data Exchange (WZDx). 2020. <https://www.transportation.gov/av/data/wzdx> Accessed May 31, 2021.